\documentclass[twocolumn]{openjournal}
\usepackage{graphicx}
\usepackage{physics}
\usepackage[colorlinks,linkcolor=blue,citecolor=blue,urlcolor=blue ]{hyperref}
\usepackage{float}
\usepackage{xcolor}
\usepackage{orcidlink}
\newcommand{\grackle}{\texttt{Grackle}}
\newcommand{\krome}{\texttt{Krome}}

\title{\grackle Paper (title pending)}

\begin{document}
\title{Bridging Machine Learning and Cosmological Simulations: Using Neural Operators to emulate Chemical Evolution \vspace{-1.5cm}}
\author{Pelle van de Bor$^{1*}$}
\author{John Brennan$^1$\orcidlink{0000-0002-4428-6798}}
\author{John A. Regan$^1$\orcidlink{0000-0001-9072-6427}}
\author{Jonathan Mackey$^2$}

\affiliation{$^1$Centre for Astrophysics and Space Science Maynooth, Department of Physics, Maynooth University, Maynooth, Ireland} 
\affiliation{$2$ Astronomy \& Astrophysics Section, School of Cosmic Physics, Dublin Institute for Advanced Studies, DIAS Dunsink Observatory,
Dublin D15 XR2R, Ireland}
\email{$^*$email: pelle.vandebor.2024@mumail.ie}

\begin{abstract}

\noindent The computational expense of solving non-equilibrium chemistry equations in astrophysical simulations poses a significant challenge, particularly in high-resolution, large-scale cosmological models. In this work, we explore the potential of machine learning, specifically Neural Operators, to emulate the \grackle{} chemistry solver, which is widely used in cosmological hydrodynamical simulations. Neural Operators offer a mesh-free, data-driven approach to approximate solutions to coupled ordinary differential equations governing chemical evolution, gas cooling, and heating. We construct and train multiple Neural Operator architectures (DeepONet variants) using a dataset derived from cosmological simulations to optimize accuracy and efficiency.

Our results demonstrate that the trained models accurately reproduce \grackle{}’s outputs with an average error of less than 0.6 dex in most cases, though deviations increase in highly dynamic chemical environments. Compared to \grackle{}, the machine learning models provide computational speedups of up to a factor of six in large-scale simulations, highlighting their potential for reducing computational bottlenecks in astrophysical modeling. However, challenges remain, particularly in iterative applications where accumulated errors can lead to numerical instability. Additionally, the performance of these machine learning models is constrained by their need for well-represented training datasets and the limited extrapolation capabilities of deep learning methods.

While promising, further development is required for Neural Operator-based emulators to be fully integrated into astrophysical simulations. Future work should focus on improving stability over iterative time steps and optimizing implementations for hardware acceleration. This study provides an initial step toward the broader adoption of machine learning approaches in astrophysical chemistry solvers.

\end{abstract}
\maketitle
\section{Introduction}
\noindent Hydrodynamical simulations of many astrophysical and cosmological processes have become a standard tool in astrophysics over the last two to three decades. Indeed, many flagship cosmological (radiation-)hydrodynamical simulations have been conducted with great success over the last decade and a half alone including \texttt{Illustris} \cite[e.g.][]{Vogelsberger_2014a, Vogelsberger_2014b, Genel_2014, Sijacki_2015}, \texttt{Renaissance} \cite[e.g.][]{Xu_2013, Chen_2014, OShea_2015, Smith_2018}, \texttt{Eagle} \citep{Schaye_2015, Crain_2015, Schaller_2015}, \texttt{Obelisk} \citep{Trebitsch_2021}, \texttt{Horizon-AGN} \citep{Dubois_2014, Dubois_2015, Dubois_2016}, \texttt{Edge} \citep{Rey_2019} and \texttt{Simba} \citep{Dave_2019} to name but a few. All of these hydrodynamical simulations have, in one form or another, had to deal with radiative cooling and chemistry. \\
\indent In fact, modelling of plasma chemistry and radiative cooling is crucial in understanding a wide range of astrophysical phenomena. Almost all astrophysical objects originate from diffuse clouds of gas and/or plasma drawn and crushed together within a deep potential well. This well is typically formed either by the plasma itself, as in the case of stars, or by a parent dark matter halo, as seen in hierarchical structure formation.

In the absence of additional physical processes, the plasma naturally settles into a state where its pressure roughly balances gravitational forces, preventing further evolution without external influence. For astrophysical structures to form, a mechanism must exist to enable energy loss, thereby breaking this equilibrium. Radiative cooling serves this essential role, often driven by a series of chemical reactions that enhance the plasma's ability to dissipate energy and ultimately enable a runaway increase in density.

Radiative cooling is a fundamental factor in several key astrophysical processes. 
In the context of cosmological structure formation, gas collapsing into dark matter halos typically encounters a strong shock near the virial radius, heating it to temperatures close to the virial temperature. When this plasma undergoes optically-thin radiative cooling, it enables the gas to condense at the centre of the dark matter halo, eventually giving rise to molecular clouds and stars \citep{Rees_1977, White_1978, White_1991}. The role of chemistry in the evolution of astrophysical plasmas is equally significant. The formation of simple molecules through gas- and dust-phase reactions can greatly enhance cooling efficiency \citep{Hollenbach_1979, Hollenbach_1989, Wolfire_1995}.\\ 
\indent In star formation, the intricate interplay between gas- and dust-phase chemistry governs the evolution of pre-stellar clouds and can significantly impact the resulting stellar initial mass function. Early models of the first stars in the Universe heavily relied on developing accurate chemistry solvers \cite[e.g.][]{Katz_1996, Abel_1997, Anninos_1997}, which were essential for hydrodynamical simulations of primordial star formation \cite[e.g.][]{Bromm_1999, Bromm_2002, Abel_2002,Turk2009a, Stacy2010a}.

\indent On smaller physical scales, radiative cooling plays a crucial role in shaping the structure and dynamics of accretion disks surrounding stars and compact objects \cite[e.g.][]{Blandford_1982, Balbus_1991}. The thermal evolution of diffuse astrophysical plasmas is largely dictated by the form of the ‘cooling curve’—which describes how the cooling rate varies with density and temperature. This cooling behaviour drives thermal instabilities, leading to the formation of a multiphase interstellar medium \citep{McKee_1977, Sutherland_1993}. It is modelling this multi-phase medium that is now often the goal of many high resolution cosmological simulations \citep[e.g.][]{Kannan_2025}. 

\indent However, chemical networks are complex and depending on the number of species involved, the number of ordinary differential equations (ODEs) that need to be solved can be significant. Moreover, the varying time scales of the different chemical reactions can result in a series of `stiff' equations, requiring computationally expensive implicit solvers to converge to a well defined solution. 
When embedded within an already complex (radiation)-hydrodynamical code this extra burden can make the overall computational footprint extremely expensive. 
While current chemistry solvers, for example \grackle{} \citep{Smith_2018}, \texttt{CHIMES} \citep{richings_2014a, richings_2014b}, and \krome{} \citep{grassi2014} are computationally optimised they nonetheless carry a significant computational cost. \\
\indent The need for methods to speed up the simulation of chemical and radiative processes warrants an investigation of potential machine learning (ML) approaches. 
ML approaches have quickly become relevant in the field of astrochemistry, opening several innovative approaches for solving ODEs by leveraging their ability to approximate complex functions in a mesh-free, data‐driven manner. Techniques such as Latent ODEs with autoencoders \citep[e.g.][]{grassi_2022, maes_2024}, Neural ODEs with autoencoders \citep{sulzer_2023}, Neural Networks \citep[e.g.][]{grassi_2011, demijolla_2019, holdship_2021, palud_2023}, Physics-Informed Neural Networks \citep{branca_2023}, Neural Fields \citep{asensio_2024}, and Neural Operators \citep{branca} promise fast and efficient approximations to solutions of chemical rate equations, with a high degree of accuracy. Benchmarking tools such as \texttt{CODES} \citep{janssen_2024} are starting to get off the ground, rounding out a developing ML ecosystem in the field. These codes are as of now standalone, with few attempts to integrate them into larger codes such as cosmological simulations. Therefore, it is natural to explore the potential of solving the computational cost problem with an ML method.


In this paper, we follow the work of \cite{branca} and investigate Neural Operators as a model for emulating the popular chemistry solver \grackle{}. We explore and report on alternative architectural choices for the Neural Operator that demonstrate an improvement in both accuracy and efficiency. We also highlight some challenges that may arise, and would need to be addressed, when integrating such an emulator into a hydro-dynamic simulation code where it would be invoked regularly.\\
The structure of the paper is as follows: In \S \ref{Sec:Methodology} we describe the methodology of both \grackle{} and our machine learning framework. In \S \ref{sec:data} we describe how we training the required datasets, in \S \ref{Sec:Results} we present our results and in \S \ref{Sec:Discussion} we discuss our conclusions.

\section{Methodology} \label{Sec:Methodology}
\subsection{Chemistry solver} \label{subsec:chemistrySolver}
\noindent The aim of this project is to create an emulator to predict solutions to the non-equilibrium chemistry equations solved by \grackle{}. \grackle{} \citep{grackle} is a stand-alone chemistry solver that can be integrated with many commonly used cosmological simulation codes [e.g. \texttt{Enzo} \citep{Bryan2014}, \texttt{Gadget} \citep{Springel_2005}, \texttt{Arepo} \citep{springel_2011}]. \grackle{} includes treatment of chemical evolution, gas heating and cooling, metallicity, dust influences, and radiation effects. \grackle{} has been used in many simulation efforts, including \texttt{SIMBA} \citep{Dave_2019}, \texttt{FOGGIE} \citep{peeples_2019}, and \texttt{AGORA} \citep{roca_2020}, and features several different chemical networks designed to model the evolution of different species. In this work, we focus on solving the 9-species network, with no radiation or metallicity influence. The species in this network are e$^-$, H$^+$, H, H$^-$, H$_2$, H$_2^+$, He, He$^+$, and He$^{++}$. Specifically, the inclusion of molecular hydrogen, H$_2$, is crucial for cosmological simulations, as molecular hydrogen rapidly cools the gas down to temperatures close to 200 K allowing for the high densities required for the first stars in the Universe \citep{Palla_1983, Abel_1998, Bromm_1999, Bromm_2002, Abel_2002}. The conditions of zero metallicity and cooling dominated by molecular hydrogen are similar to that in the early Universe, before the formation of metal-producing stars \citep[e.g.][]{Bromm_2004, Yoshida_2006,Bromm_2009}. 

The chemical network that \grackle{} uses for the nine-species model, without radiation and dust,  consists of 23 chemical reactions, combined into ten rate equations, one for each species and one for the internal energy. The chemistry solver works on uniform regions of space, typically the smallest subdivisions of space considered by hydrodynamic simulations (either cells or smoothed particle hydrodynamic (SPH) particles) - we will use the term cell going forward to refer to input data encompassing both species densities and energies. \grackle{} updates the species abundances and temperature for each cell locally. The rate equations take the general form
\begin{equation}
    \pdv{n_i}{t} = \sum_j\sum_lk_{jl}(T)n_jn_l + \sum_jI_jn_j
\end{equation}
where $n_x$ represents the number density of species $x$, $k_{jl}$ is the reaction rate between species $j$ and $l$, which is dependent on the temperature $T$, and $I_j$ the radiative rate. A positive $k_{jl}$ indicates a creation term, while a negative $k_{jl}$ indicates a destruction term. As a constraint, the H$_2^+$ species is always assumed to be at equilibrium since the reaction timescale for this species is short enough to be decoupled from the main integration. Since radiation is not included in this work, $I_j=0$ for all species. We refer the reader to \citet{grackle} for further details on the rate equations that are included in \grackle{}'s nine species network, as well as the values of $k_{jl}(T)$. \\

To solve these equations, \grackle{} solves each equation in time until a final time $t_f$, usually the hydrodynamical time step provided by the simulation, has been reached. These rate equations form a coupled ODE system, for which the solution can be written as an operator $\mathcal{G}$ acting upon the time step $t_f$, the species densities, and the internal energy: 
\begin{equation} \label{eq:operator}
    n_i(t_f), E(t_f) = \mathcal{G}(n_i(0), E(0))(t_f).
\end{equation}

\grackle{} also performs heating and cooling operations, updating the local temperature. One point to note is that \grackle{} works directly with the internal energy of the gas instead of the temperature of the gas, as this is a more common parameter to be used by simulation codes. The conversion between temperature and internal energy follows the formula 
\begin{equation}
    E = \frac{kT}{(\gamma-1)\mu m_\mathrm{H}},
\end{equation}
where $E$ is the internal energy (in erg/g), $k$ the Boltzmann constant, $T$ the gas temperature, $\gamma$ the adiabatic index of an ideal gas (in this work assumed to be $5/3$), $\mu$ the mean molecular weight of the gas as a fraction of hydrogen mass, and $m_\mathrm{H}$ is the hydrogen mass.
Updating the internal energy is done according to 
\begin{equation}
    \dv{e}{t}=-\Dot{e}_\mathrm{cool}+\Dot{e}_\mathrm{heat}
\end{equation} where $\Dot{e}_\mathrm{cool}$ represents cooling terms, and $\Dot{e}_\mathrm{heat}$ represents heating terms. Selecting which cooling effects to take into account is highly customizable for \grackle{}. We use the non-equilibrium primordial heating rates for the nine species model as outlined in \cite{grackle}. This consists of collisional excitation, collisional ionization, recombination cooling, brehmsstrahlung, Compton heating/cooling off the CMB, photoionization heating, and H$_2$ cooling. Further details on the rates used by \grackle{} can be found in \cite{grackle}.
\\
\indent Since all the species evolve at the same time, \grackle{} uses a solver based on the backwards difference formula method, with alterations through partial updates and subcycling, following \cite{Anninos_1997}. For large values of $t_f$, relative to the reaction timescale of the species, the relative change for each species can be excessive, which can cause numerical errors in the integrator. The subcycling method divides $t_f$ into smaller intervals $\tau$ such that no changes larger than 10 \% per subcycle occur for H, e$^-$, and the internal energy. \\
\indent Since this only evolves the rate equations for a fraction of $t_f$, this method necessitates calling the solver several times. This increases both the numerical stability and the accuracy, but also significantly increases the computational load - this can be particularly troublesome in high density, star-forming, regions. This computational load is a primary bottleneck in large scale, high resolution, cosmological simulations. Decreasing this computational load, while retaining the accuracy is a potentially significant and perhaps necessary feature for future high-volume, high-resolution, simulations. One proposed method of achieving this is through machine learning.

\subsection{Neural Operators}
A major development in the theoretical understanding of neural networks was the Universal Approximation Theorem (\cite{cybenko1989approximation}, \cite{hornik1990universal}), which states that neural networks can approximate a large class of functions to an arbitrary level of precision. Interestingly, this theorem can be generalized to state that a modified neural network, known as a neural operator, can similarly approximate operators \citep{chen1993approximations, deeponet}. Neural operators have proven effective in learning complex phenomena, including multiphase flow \citep{Wen_2022}, plasma modelling \citep{Gopakumar_2023}, and identifying attractors in chaotic systems \citep{Li_2022}.

\begin{figure}
    \centering
    \includegraphics[width=\linewidth]{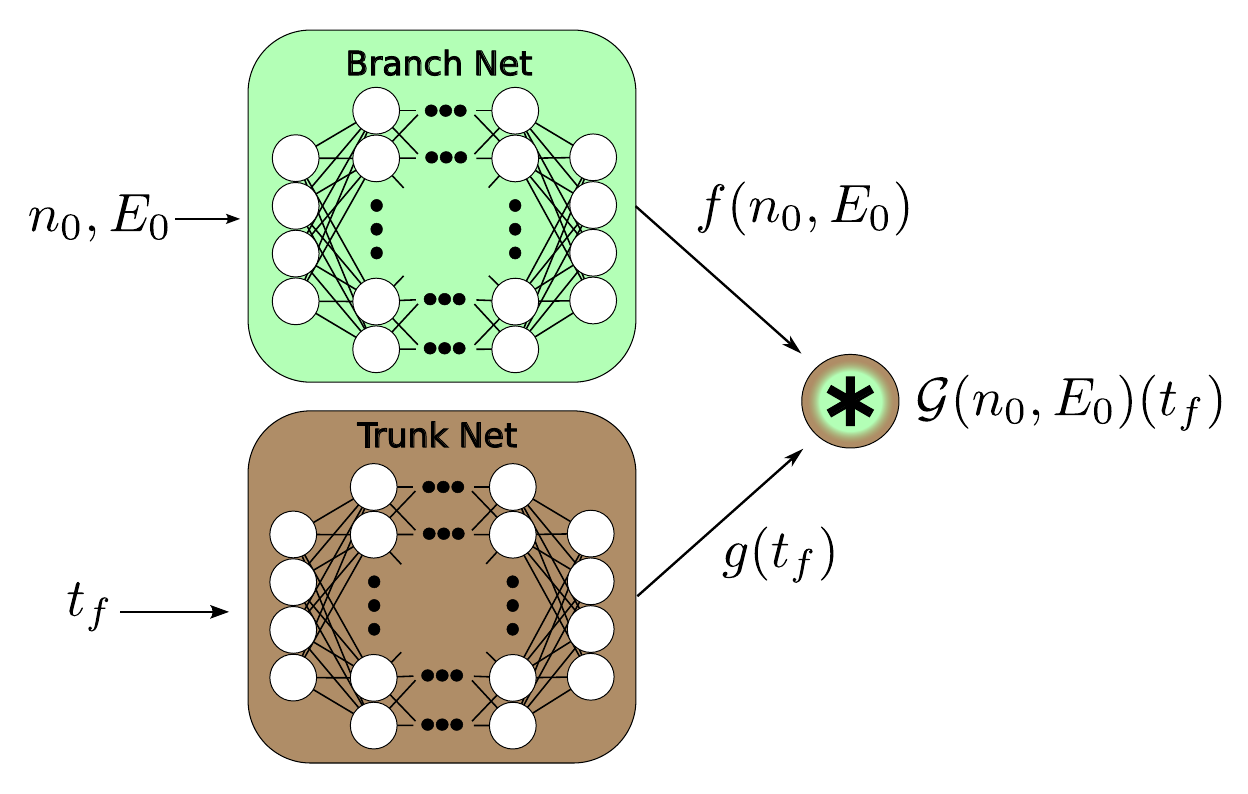}
    \caption{A visualization of the DeepONet high-level architecture as used here. The input variables $n_0$, $E_0$, and $t_f$ get transformed through their respective networks, the outputs of which are combined using the dot product to create the output of the DeepONet.\\}
    \label{fig:deeponet}
\end{figure}

One significant application of neural operators lies in solving coupled ODEs, such as the chemical rate equations. This was demonstrated in the work of \cite{branca} who employed a model, known as DeepONet, to emulate the \krome{} chemistry solver \citep{grassi2014}. DeepONet \citep{deeponet} (Deep Operator Network) is a high-level neural operator architecture consisting of two fully connected neural networks (FCNNs) whose outputs are combined to generate the final output. In our configuration, one of the FCNNs, known as the branch network, is designed to take an 11-dimensional vector as input. The components of this vector consist of the densities of the nine species $n_i(0)$, the total density of the gas $n(0)$ and the internal energy $E(0)$. These determine the initial conditions and some of the parameters (namely the reaction rates) of the rate equations. We include the total density as a sanity check. The chemistry solver does not evolve the total density, therefore the lack of evolution of this parameter may serve as a first test on convergence. The second FCNN, known as the trunk network, is designed to take a single number as input: the time, $t_f$, over which the initial conditions are evolved. Both FCNNs produce vectors of configurable size, and DeepONet computes its final output by taking the dot product of these vectors. A visualization of the structure as used in this work is shown in Figure \ref{fig:deeponet}.

The standard DeepONet architecture,  just described, outputs a single number. However, since the state of the system in this context consists of ten densities and an internal energy, to learn the solution operator in Equation \ref{eq:operator}, a model that can output an 11-dimensional vector is required. There are several strategies for extending the DeepONet to multiple outputs, as proposed by \citet{deeponetMulti}. The most straightforward is to train separate, independent DeepONets for each parameter, but this is computationally expensive. It is also possible to split the output vector of the branch network into $N$ smaller vectors and then perform the dot product between each of these vectors and the trunk output, producing $N$ outputs. This requires that the size of the branch network output is $N$ times that of the last layer of the trunk network. For this work, $N=11$, as there are 11 predicted parameters. The first method will be referred to as the \texttt{Independent} strategy, and the second as \texttt{branchSplit} strategy. \citet{deeponetMulti} propose more alternate methods which are beyond the scope of this work.

\subsection{Model Variants}
The fiducial model we present here has the same network architecture as proposed by \cite{branca}. While this particular choice of architecture has been shown to be reasonably accurate, we look to conduct a parameter study to explore alternatives. To this end, we introduce three additional models besides the fiducial model, which we refer to as \texttt{Wide}, \texttt{Deep}, and \texttt{WideDeep}. These models are variations on the fiducial model, changing the number of layers, nodes per layer, and multi-output strategy. The \texttt{Wide} model is both shallower including fewer layers) and wider (having more nodes per layer) than the fiducial model, the \texttt{Deep} model has a deeper network (including more layers), while \texttt{WideDeep} is both deeper and wider than the fiducial model. The details of these models are shown in Table \ref{tab:models}. The \texttt{Wide} model has fewer parameters than the fiducial model, to find a more lightweight model capable of emulating the chemical network, as well as determining the influence of the \texttt{branchSplit} output strategy. The \texttt{Deep} model is focused on depth, to probe whether deep learning techniques may be a requirement for future studies. Finally, the \texttt{WideDeep} model is meant to see if a large quantity of parameters can provide an alternative in the accuracy/speedup trade-off. 
For these models, the shape of the network is identical between branch and trunk networks, with one exception. The final layer of the branch net for models using the \texttt{branchSplit} output strategy has 11 times as many nodes as the last layer of the trunk net, since our network features 11 unique outputs (the densities for the nine species, the energy, and the total density). The final layer of the trunk net of the \texttt{Wide} model has 128 nodes, meaning that the branch net has $11\cdot128=1408$ nodes. The output layers of the branch and trunk networks are combined using the dot product to generate 11 output variables. 
For all the models, the branch network also features an input layer with 11 nodes, while the trunk network features an input layer with 1 node. This input layer is not shown in Table \ref{tab:models}.

\begin{table}[]
    \centering
    \begin{tabular}{c||c|c|c}
        Model & Layers & Nodes per layer & Output strategy \\ \hline
        Fiducial & 6 & 128 & \texttt{Independent}\\
        \texttt{Wide} & 3 & 1024$^*$ & \texttt{branchSplit}\\
        \texttt{Deep} & 16 & 64 & \texttt{Independent}\\
        \texttt{WideDeep} & 16 & 512$^*$ & \texttt{branchSplit} \\
    \end{tabular}
    \caption{An overview of the networks for the proposed ML models. All models have an input layer, consisting of 1 node for the trunk net and 11 nodes for the branch net.\\
    $^*$The last layer of the \texttt{Wide} net features 128 nodes for the trunk net, and 1408 nodes for the branch net, while the last layer of the \texttt{WideDeep} branch net has 5632 nodes.}
    \label{tab:models}
\end{table}

\section{Data and Training}\label{sec:data}
To construct a meaningfully large data set, appropriate values for the energy, the time, $t_f$, and each density must be found. One can set arbitrary minimum and maximum values for these quantities, but this comes with a pitfall: 
DeepONet models are known to be poor at extrapolating outside the ranges covered by the training set. For example, \cite{branca} find that this is especially true for the time domain. It is therefore important to select the appropriate range of densities, energies, and times, to capture the behaviour that one wishes to emulate. 

\subsection{Cosmological Simulation}
\noindent Since the chemical network we aim to emulate excludes metals and radiation, the gas state we model and train on reflects the early Universe's chemistry—before significant star formation begins. To determine the necessary range of densities and temperatures for our dataset, we conducted a cosmological simulation using Grackle as the chemistry solver. We then analyzed the simulation outputs to extract realistic parameter ranges, which we used to define the dataset boundaries.

Our primary objective in this simulation was to achieve high spatial resolution, allowing us to capture a diverse range of environments, from low- to high-density regions. This diversity broadens the dataset’s coverage, enhancing the robustness of the trained model and improving its ability to generalize across different astrophysical conditions.

\begin{figure}
    \centering
    \includegraphics[width=\linewidth]{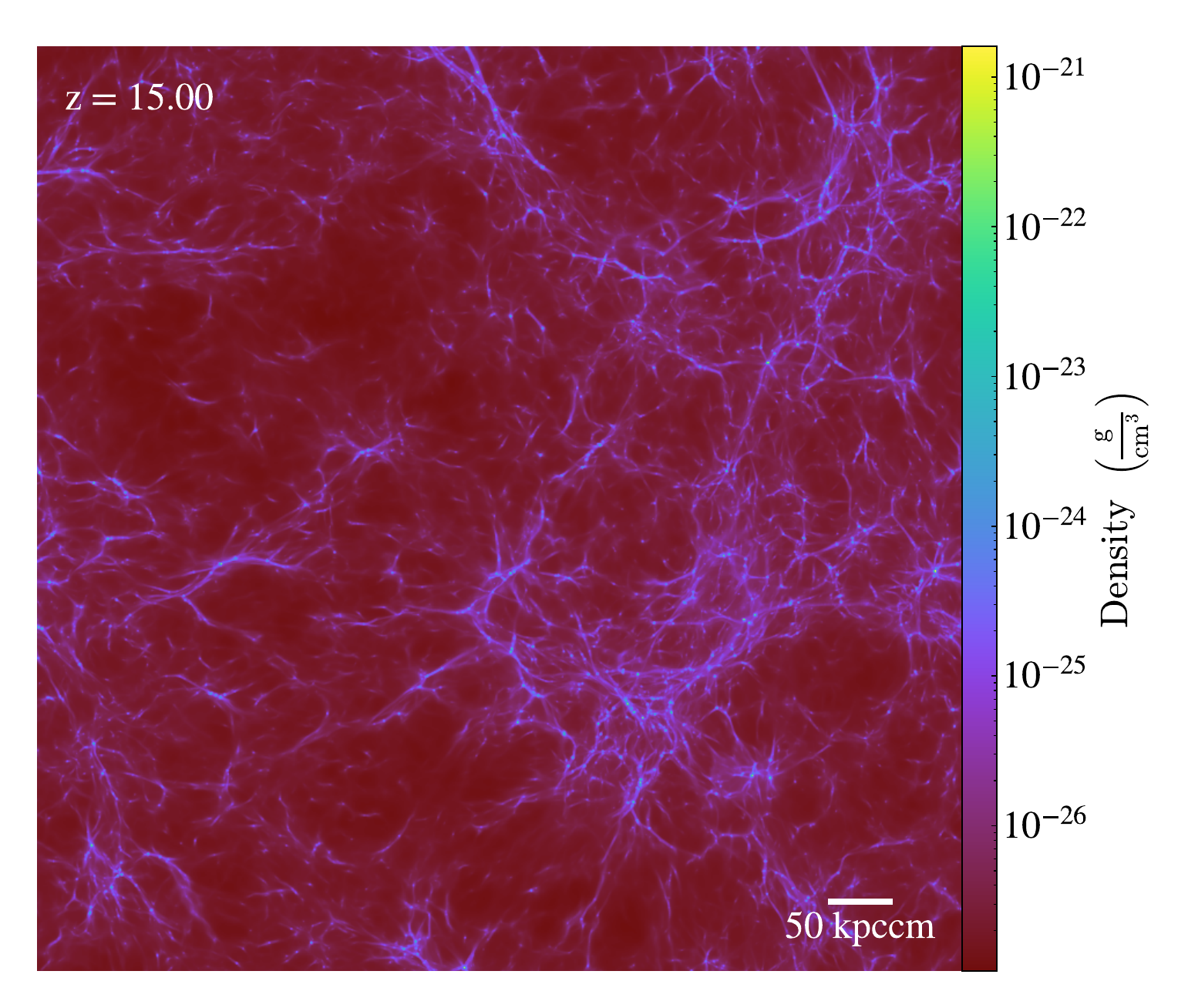}
    \caption{Visualization of a slice through the centre of the density field of the cosmological simulation used in this work at $z=15$. This simulation has high spatial resolution (16 cpc) to increase the density variance in a small region. The density spans over 5 orders of magnitude in just this slice.}
    \label{fig:simulation}
\end{figure}

\indent Initial conditions for the simulation were created using MUSIC \citep{Hahn2011}, and evolved using the \textsc{ENZO} cosmological simulation code \citep{Bryan2014} down to $z=15$. The initial conditions consist of a $0.5$ cMpc$/h$ box, consisting of $512^3$ cells at $z=127$, with a maximum refinement level of 6. The minimum cell length is therefore $15$ cpc$/h$. This small scale allows gas to take on a wide range of densities. The minimum particle mass is 99 M$_\odot$. For this simulation, we assume a flat cosmology with parameters $\Omega_\texttt{b}=0.0487$, $\Omega_\texttt{c}=0.2602$, $\Omega_\Lambda=0.6911$, and $h=0.6774$, according to \cite{planck2020}. For the chemical network used in the simulation, we use the nine species chemical network as described in \S \ref{Sec:Methodology}, which excludes metals, dust, and radiation.
A view of the simulation results at $z=15$ is shown in Figure \ref{fig:simulation}, showing that the density ranges from $10^{-21}$ g/cm$^{3}$ to $10^{-26}$ g/cm$^{3}$ in this view.
To determine the range of values for each species to use to create a data set, we use the minimum and maximum value present in the simulation. In order to broaden the ranges, since this single simulation may, through chance, be a rare region, these minimum and maximum values are rounded down/up to the nearest integer in logarithmic space, respectively. This captures the entire simulation within the used ranges, and allows for possible densities in rarer situations to also be captured by the trained model. For $t_f$, we extract the values used as input into \grackle{} at runtime. This is typically the hydrodynamical time step.
We elect to round down the highest value for $t_f$, since the DeepONet model is more sensitive to changes of $t_f$ than any other parameter. Values of $t_f>10^6$ years are also extremely rare (for instances where a non-equilibrium solver is required), with this maximum value being an outlier. The ranges for each parameter found and the ranges used for our data set are shown in Table \ref{tab:ranges}. We base our ranges off of mass densities $\rho_\mathrm{i}$, and convert them to the appropriate units for \grackle{}. \\

\subsection{Training}
\noindent To generate a training dataset, we generate random samples of inputs by uniformly sampling each parameter in logarithmic space independently, sampling the entire parameter space uniformly. For each parameter $x$, the minimum and maximum values ($x_\mathrm{min}$ and $x_\mathrm{max}$) are chosen from the ranges specified in Table \ref{tab:ranges}. For each input $((n_j, E), t_f)$, associated outputs $G((n_j, E), t_f)$ are computed using \grackle{}, forming an input-output pair, also referred to as a sample. The samples are normalized by rescaling the values between -1 and 1, using \begin{equation}
    x_{i\mathrm{,norm}}=2\frac{x_i-x_{i\mathrm{,min}}}{x_{i\mathrm{,max}}-x_{i\mathrm{,min}}} - 1
\end{equation} following \cite{branca}. The training set consists of $10^7$ of such normalized samples, while the test set, since it is only used to validate the model, can be smaller than the training set. It therefore consists of $10^6$ samples. We chose a training set of size $10^7$ to obtain a population of samples over the entire parameter space, while keeping memory and computational constraints into account. It is important to note that these samples may have non-chemical parameter combinations. That is to say, charge is not conserved and the H/He ratio is not identical in each sample. We elected to use this method to allow for more degrees of freedom for the prediction, and to capture the performance of \grackle{} over the entire parameter space. We discuss this point further in \S {\ref{sec:iteration}.} \\
\indent To train the models, we use the \texttt{DeepXDE} package, which is designed to train DeepONet models. This package is backend-agnostic, we select a \texttt{Tensorflow} backend. All models were initialized using the Glorot initialization \citep{glorot2010understanding}, use the ReLU activation function, and are evolved with a learning rate of $10^{-3}$ using the Adam optimizer \citep{kingma2014adam} for $4\cross10^4$ iterations, until the fiducial model showed no major improvements on the validation set. To prevent potential overfitting, each model uses L1 and L2 regularization. This training procedure follows the procedure laid out by \cite{branca}. Training was conducted on two AMD EPYC 7452 32-Core CPU chips, with each model taking up to 4 days to complete training. These chips were chosen due to hardware constraints. \\

\begin{table}
    \centering
    \begin{tabular}{c|cc|cc|c}
        Parameter & Min & Max & Range Min & Range Max & Unit \\\hline
        $\rho$ & -27.6 & -19.1& -28 & -19 & log(g/cm$^3$) \\
        $\rho_\mathrm{H}$ & -27.7& -19.2& -28 & -19 & log(g/cm$^3$)\\
        $\rho_\mathrm{H^+}$ & -31.5& -25.1& -32 & -25 & log(g/cm$^3$)\\
        $\rho_\mathrm{H^-}$ & -40.3 & -31.2& -41 & -31 &log(g/cm$^3$)\\
        $\rho_\mathrm{H_2}$ & -32.9& -21.5&-33 & -21 & log(g/cm$^3$)\\
        $\rho_\mathrm{H_2^+}$ & -41.7& -29.2&-42 & -29 & log(g/cm$^3$)\\
        $\rho_\mathrm{He}$ & -28.2& -19.7&-29 & -19 & log(g/cm$^3$)\\
        $\rho_\mathrm{He^+}$ & -41.7& -29.2&-42 & -29 & log(g/cm$^3$)\\
        $\rho_\mathrm{He^{++}}$ & -46.7& -40.3&-47 & -40 & log(g/cm$^3$)\\
        $\rho_\mathrm{e^-}$ & -31.5& -25.1& -32 & -25 & log(g/cm$^3$)\\
        $E$ & 8.6 & 12.9& 8 & 13 & log(erg/g)\\
        $t_f$ & -2.6& 6.3& -2 & 6 & log(yr)
    \end{tabular}
    \caption{Minimum and maximum values for each parameter as found in the simulation, as well as the selected minimum and maximum values used in creating the training and test data sets, based on the values present in the cosmological simulation. The 9 species mass densities are present, as well as the total density, the internal energy of the gas, and the external time step.}
    \label{tab:ranges}
\end{table}

\section{Results} \label{Sec:Results}
\noindent To test our neural operator network we perform several tests between \grackle{} and the ML predictions. The first test aims to validate the evolution of a single cell over the entire time domain. A second test compares the results of evolving a large quantity of samples, and provides the relative error between \grackle{} and the ML models. Finally, we present a direct comparison of the computational performance between \grackle{} and the ML models.

\subsection{One Zone Test} \label{sec:onezone}
\begin{figure}
    \centering
    \includegraphics[width=0.4\textwidth]{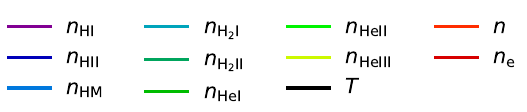}
    \includegraphics[width=\linewidth]{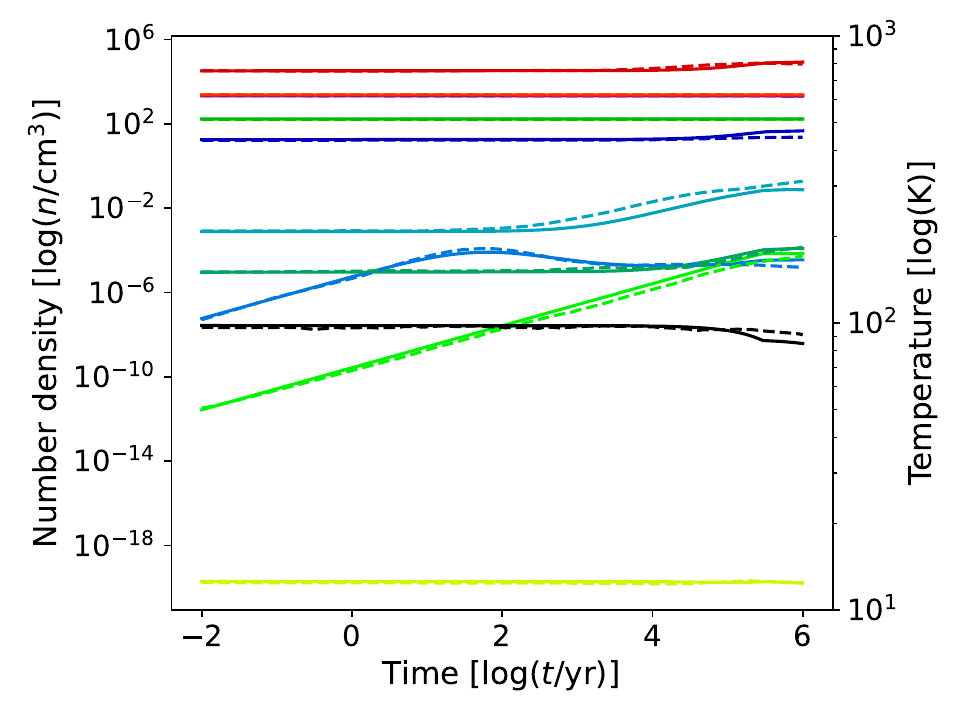}
    \includegraphics[width=\linewidth]{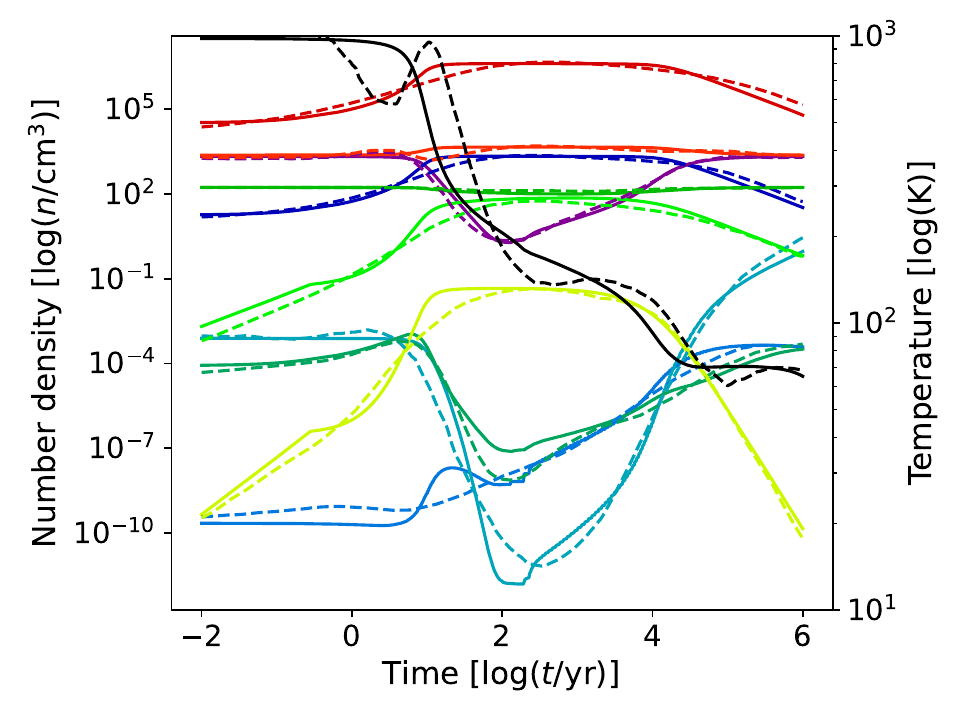}
    \caption{One zone test based on two initial cells. The solid lines show the calculated values for each species according to \grackle{}, while the dashed lines show the prediction made by the fiducial model. The top panel shows the evolution of a typical cell. The ML prediction and \grackle{} agree for the evolution of this cell up to a variation of at most 0.6 dex. The bottom panel targets a cell with a higher variance over its evolution. The disagreement between ML and \grackle{} in the bottom panel is greater, at 1.5 dex.}
    \label{fig:onezone}
\end{figure}

\noindent To show the stability and consistency of the trained fiducial model, we perform a test on one cell (one zone) and compute $G(n_0, E_0)(t_f)$ for 1000 values of $t_f$. We perform this from the same initial $G(n_0, E_0)$, randomly selecting a cell for the initial values. $t_f$ ranges from $10^{-2}$ yr to $10^6$ yr, linearly distributed in log space, following the ranges found for $t_f$ based on a single hydrodynamics time step. This forms 1000 inputs for the DeepONet. The resulting number densities and temperature predicted by this test are shown in Figure \ref{fig:onezone}. The solid lines corresponds to values calculated using \grackle{}, while the dotted lines represents the predictions made by the fiducial model. The discrepancy between \grackle{} and the fiducial model for all species in this example is below 0.6 dex. There does not appear to be a clear pattern between $t_f$ and the error. This cell, being typical of the overall population, has many species near equilibrium, which are not rapidly changing. \\
\indent The bottom panel of Figure \ref{fig:onezone} shows another example, this time with initial conditions randomly selected from the set of cells featuring $>1$ dex error for any individual species, to show a larger change in densities with highly non-equilibrium initial conditions. Here, it's clear that the DeepONet prediction struggles to capture finer details in the evolution, representing rapid changes, with errors ranging as high as 1.5 dex. This corroborates a similar result found by \cite{branca}. Cells with a rapid evolution may be rare, but their complex evolution is computationally expensive. Nonetheless, capturing their behaviour accurately will be a challenge for ML methods.

\subsection{Accuracy Test}
\begin{figure}
    \centering
    \includegraphics[width=\linewidth]{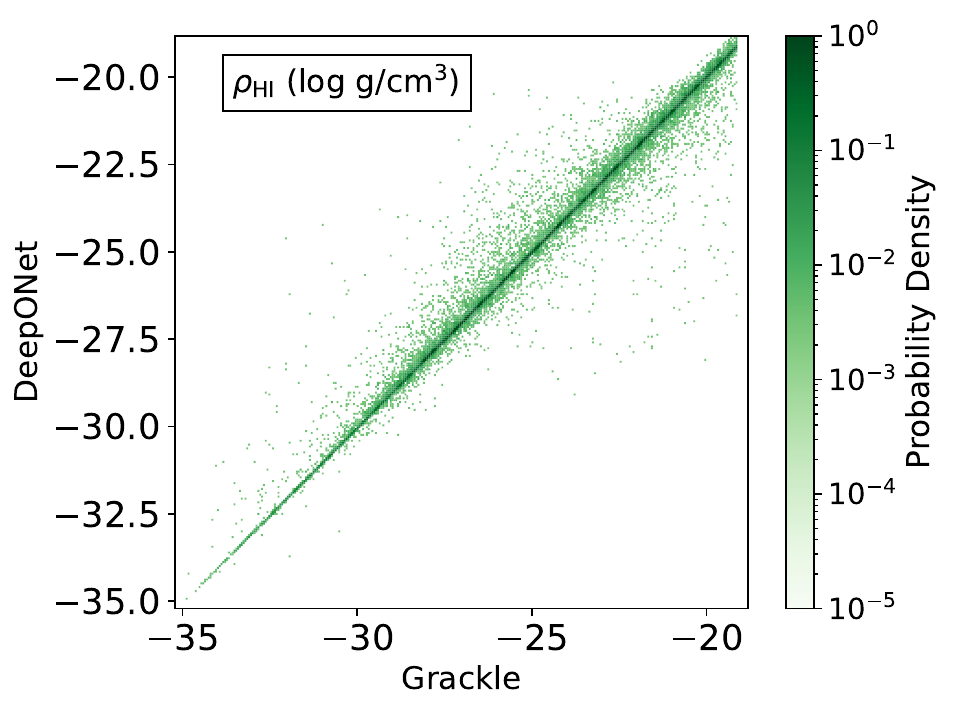}
    \caption{Probability distribution of values calculated with \grackle{} versus predictions by the fiducial model for neutral Hydrogen, based on a set of $10^6$ samples. The diagonal line represents an ideal match between \grackle{} and the fiducial model. \\}
    \label{fig:accuracy}
\end{figure}

The One Zone test shows excellent correspondence between \grackle{} and the fiducial model over the entire time domain for a single region. However, it does not test the model for a large range of initial densities and temperatures. For this, we use a different test. This test is a direct comparison of 10$^6$ samples, evolved using both \grackle{} and the fiducial model. An ideal ML model would have perfect one-to-one agreement between its predictions and \grackle{}'s calculations. A direct comparison is given in Figure \ref{fig:accuracy}, where the calculated values by \grackle{} are shown on the x-axis, while the predicted values by the fiducial model for neutral Hydrogen are shown on the y-axis. In the ideal scenario, all points would lie on a diagonal line with slope one in Figure \ref{fig:accuracy}. We can see that for the vast majority of cells, the ML prediction matches the calculations by \grackle{}, however there is still some scatter present. A direct comparison for the other species and the temperature is shown in Appendix \ref{sec:appendixAccuracy}. For these species, \grackle{} and the fiducial model similarly agree, with some species experiencing more scatter than others. The total density is matched the greatest, showing that the lack of total density evolution by \grackle{} is matched by the fiducial model. The strongest scatter is present in the temperature, but this is still relatively minor.

\subsubsection{Relative Error}
\indent To quantify the error between the previous results, we use the relative error \begin{equation}
    \Delta_r = \abs{\frac{x_G-x_i}{x_G}}
    \label{eq:relerr}
\end{equation}
where $x_G$ is the value calculated by \grackle{}, and $x_i$ the value predicted by ML. Both values are in logarithmic mass density units, as shown in Table \ref{tab:ranges}. This error can be calculated for each individual species, as well as for the temperature. Since this measure can be quantified quickly for each model, we calculate the relative error for both the fiducial model and the other models considered (i.e. for the \texttt{Deep}, \texttt{Wide} and \texttt{WideDeep} models). This allows for a direct comparison of the performance of each of these models. The distribution of the relative errors for neutral Hydrogen is shown in Figure \ref{fig:modelperformance}. The distributions of the other species and the temperature can be found in Appendix \ref{sec:appendixRelErr}. From these figures, we find that the \texttt{Wide} model has the lowest relative error over most of the species, in general matching or improving on the fiducial model, only performing worse to a significant degree on the internal energy. The \texttt{Deep} model matches the performance of the fiducial models for the internal energy and several of the species, however for species such as HI and HeI, the \texttt{Deep} model the performance is worse than either the fiducial or \texttt{Wide} model. The \texttt{WideDeep} model consistently performs worse than the fiducial model, as well as the \texttt{Wide} and \texttt{Deep} models. One possible reason for this is the identical training for the models. Models with more parameters take more training to converge, and the model performance lines up with the number of parameters in each model. Each model received an equal amount of training, for which every model has converged. This was chosen due to computational resources, and to provide a fair comparison between models. While all models have converged, it is possible that an increase in training may lead to further performance gains. We will investigate this in a future study. 

\begin{figure}
    \centering
    \includegraphics[width=0.5\textwidth]{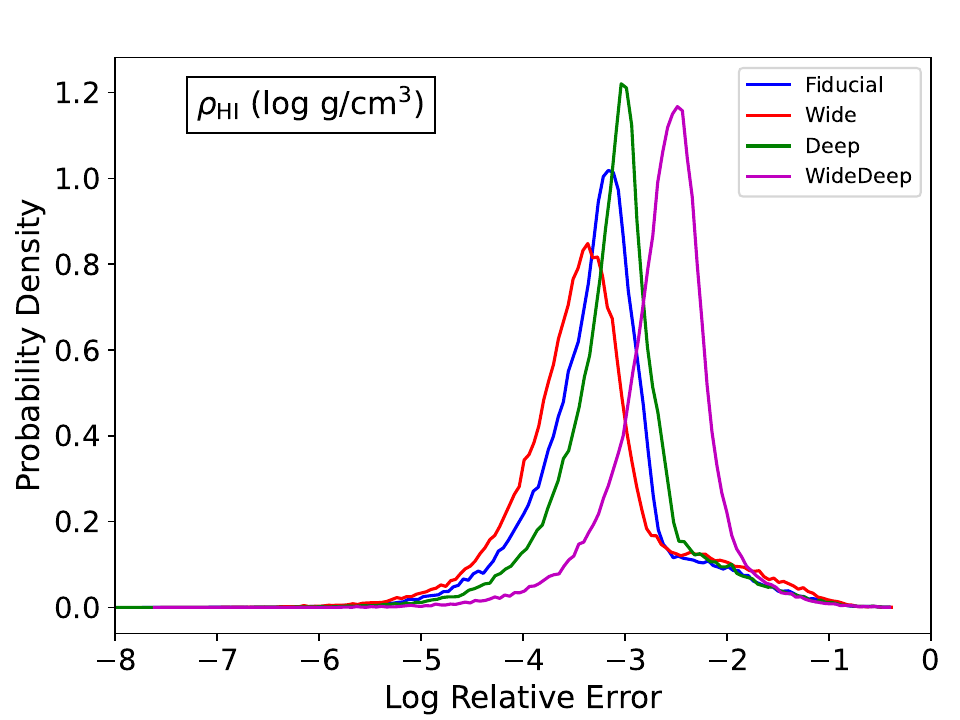}
    \caption{Probability distribution of relative errors between \grackle{} and proposed ML models for neutral Hydrogen. These values are calculated using equation \ref{eq:relerr}, based on \grackle{} and ML predictions of $10^6$ samples. }
    \label{fig:modelperformance}
\end{figure}

\subsection{Time Comparison}
\noindent A primary motivation for using an ML model over \grackle{} is the computational cost of \grackle{} in high resolution (high density) cosmological simulations. Often, the chemistry solver is the primary computational bottleneck in high resolution hydrodynamical simulations, especially in non-equilibrium regions. Meanwhile, the primary computation in neural networks consists of a series of matrix multiplications, where the computational time increases with the number of layers and nodes. The computational time is independent of the input parameters, and therefore should scale better with the number of cells considered. 
To estimate the speedup by using our ML methods, we perform a direct comparison between the two methods on the same hardware. The comparison is done by generating a set of new cells according to the previously described routines, and one value for the time step, which is used by all cells. This is similar to a grid of cells typically found in hydrodynamical simulations, where one spatial region, consisting of many cells, is evolved for a constant amount of time. \grackle{} is optimized to handle evolution on a grid like this, and therefore this should produce the most optimal performance for \grackle{}. 

To compare the scaling of the two methods, several sets of new cells are generated, with different sizes ranging from $10^3$ cells to $10^6$ cells. We measure the time to completion for each of these sets for \grackle{} and ML, and repeat this process 20 times to find an average speed, each time with newly initialized cells. We only measure the time to completion of the call functions of \grackle{} and the ML models, minimizing potential overheads. 
Computation is done on a single CPU core for consistency. This disadvantages the ML method, which is optimized for GPU computations. However, most hydrodynamical codes run exclusively on CPUs, so using this hardware gives a more practical comparison between the methods. Future optimisations may include porting chemistry solver calculations to the accelerators on chips. ML calculations are relatively easy to port over to GPUs. This can further increase speedup through parallelization. A traditional solver has to be redesigned for this hardware from scratch, with developments already underway \citep[e.g.][]{balos_2024}.

\begin{figure}
    \centering
    \includegraphics[width=0.5\textwidth]{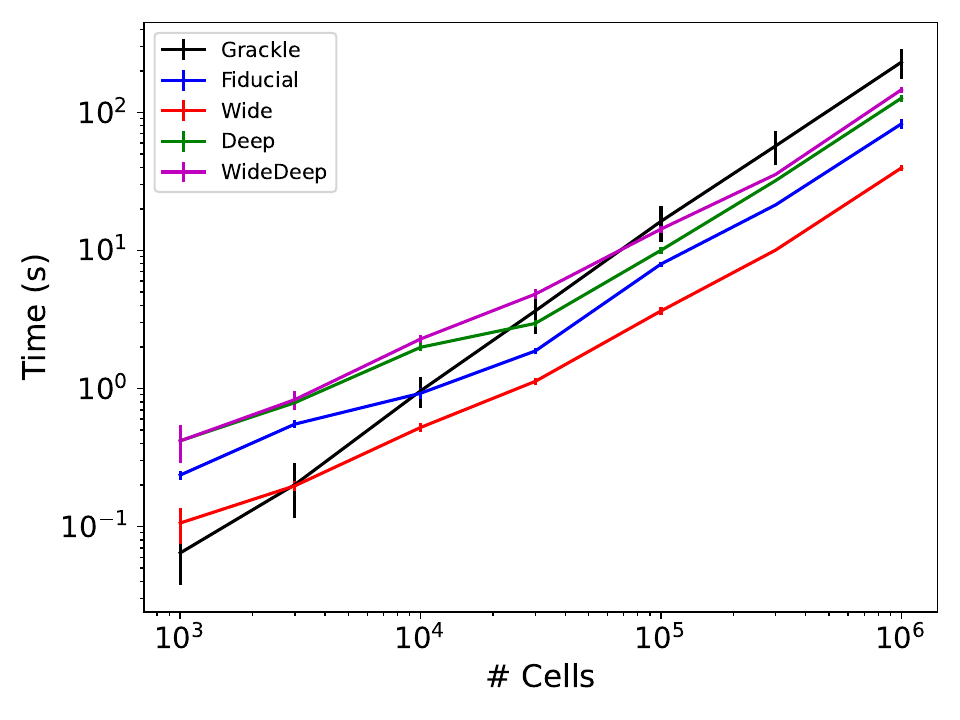}
    \includegraphics[width=0.5\textwidth]{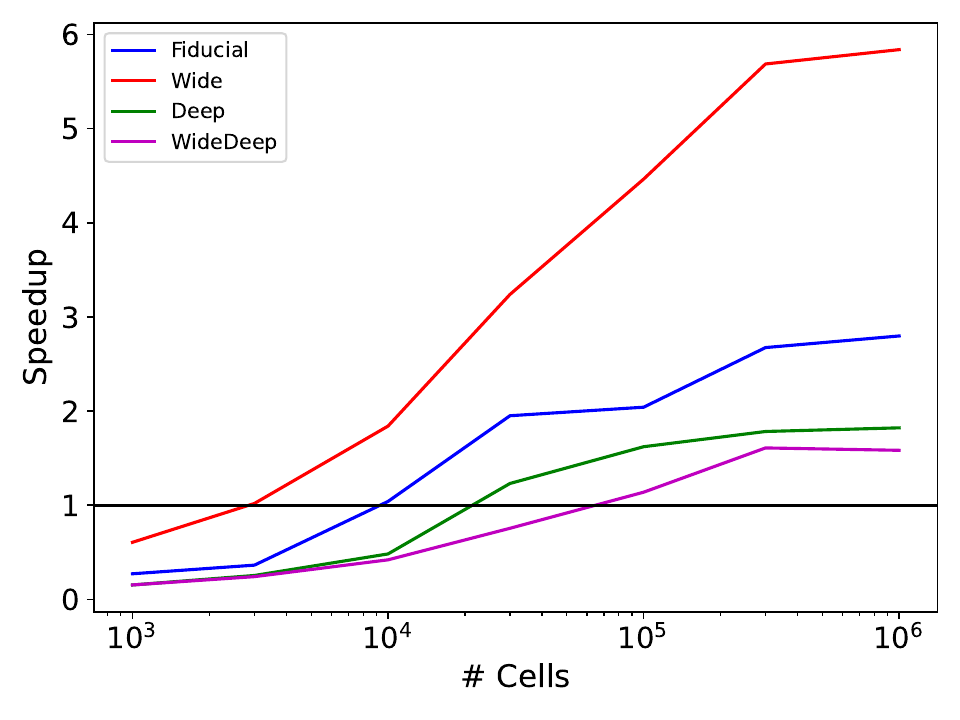}
    \caption{Computational performance comparison between \grackle{} and proposed ML models.
    Top: Comparison between mean computation time of \grackle{} and ML as a function of number of cells. The means are calculated over 20 runs for each density.
    Bottom: The speedup of ML compared to \grackle{}, calculated using the ratio of the two means, with the equal speed highlighted.}
    \label{fig:modelscompared}
\end{figure}

The resulting computational time and speedups compared to \grackle{} as a function of the number of cells are shown in Figure \ref{fig:modelscompared}. The \texttt{Wide} model shows the best speedup, which is expected since it features the fewest amount of layers. The \texttt{Deep} and \texttt{WideDeep} models show worse speedups, with \texttt{WideDeep} being the slowest model. This shows that the possible advantage of the \texttt{BranchSplit} output strategy over the \texttt{Independent} strategy appears smaller than the negative effect of the use of a wider network on computational time.
Notable is that none of the models outperform \grackle{} for $10^3$ cells. This implies that while the ML models scale better to higher cell numbers, \grackle{} is more lightweight and is a better fit for smaller cell counts. 

\begin{figure}
    \centering
    \includegraphics[width=0.5\textwidth]{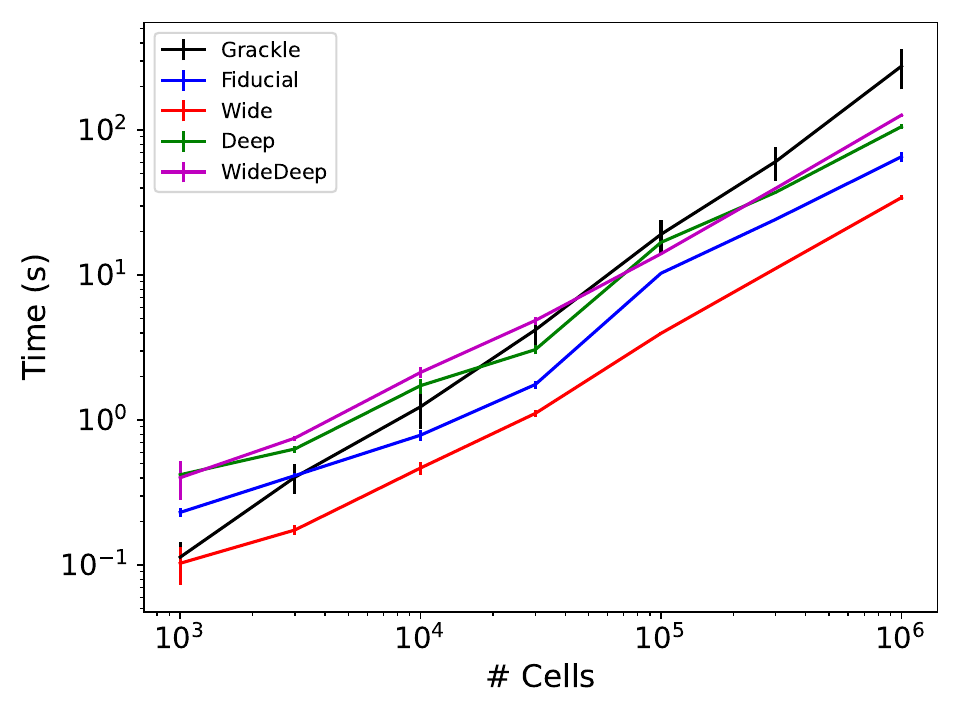}
    \includegraphics[width=0.5\textwidth]{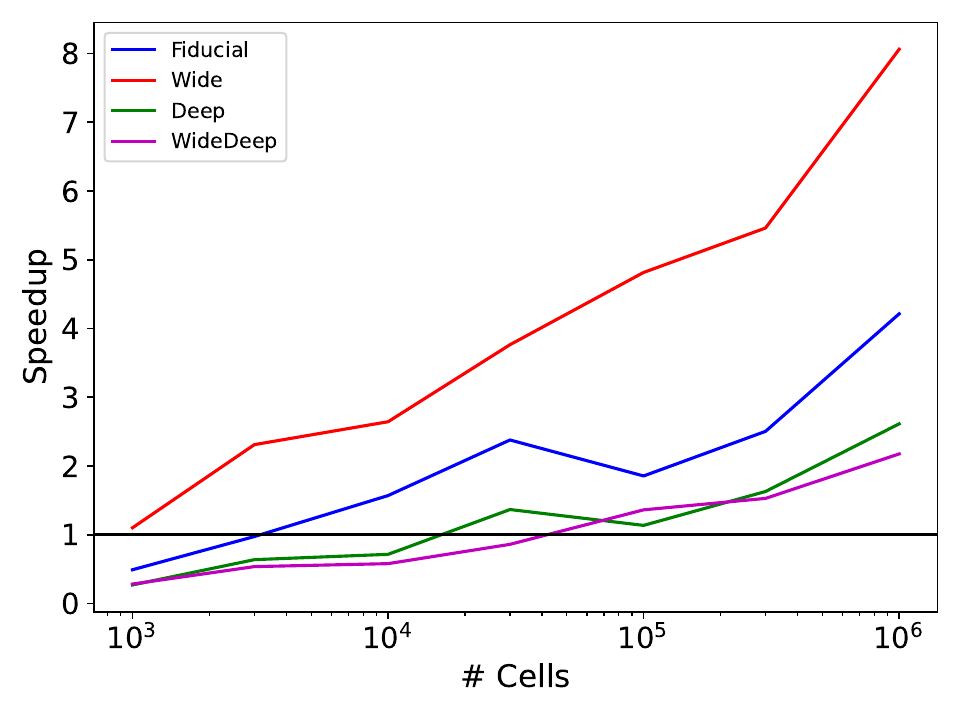}
    \caption{Top: Comparison between mean computation time of \grackle{} and the ML models as a function of number of cells. The means are calculated over 20 runs for each density.
    Bottom: The speedup of the ML models compared to \grackle{}, calculated using the ratio of the two means, with an equal speed highlighted.
    As opposed to Figure \ref{fig:modelscompared}, these times were calculated using initial values from Table \ref{tab:highdens}, as opposed to those from Table \ref{tab:ranges}.}
    \label{fig:timeshighdens}
\end{figure}

The resulting times to completion and relative speedups are shown in Figure \ref{fig:modelscompared}. These results show that ML has significant benefits over \grackle{} when comparing computation times, but it is dependent on the number of cells in the calculation. \grackle{}'s speed in equilibrium scenarios (i.e. low to medium density) is higher compared to the ML algorithm, while for non-equilibrium (i.e. high density) cells \grackle{} takes a significantly longer time to compute. A smaller number of cells will have fewer outliers in the non-equilibrium regime, which dominate the compute time of the full dataset. This explains why for a large dataset, \grackle{} performs worse in comparison to ML. \\
Therefore, we run the same test, but with different cell generation. Non-equilibrium situations play a more important role in high density regions \citep{richings_2014b, prole2024}. Therefore, a different set of initial ranges is used, with focus on high densities (i.e. star forming regions). The exact values used are shown in Table \ref{tab:highdens}. The results from this test are shown in Figure \ref{fig:timeshighdens}. Comparing these results to those shown in Figure \ref{fig:modelscompared}, we see a speed-up of the \texttt{Wide} model of a factor of up to 8 compared to \grackle{}. We note that these non-equilibrium cells, as shown in \S \ref{sec:onezone} do generally produce less accurate results. This tradeoff will have to be taken into account. A potential application of this phenomenon would use a bespoke algorithm to detect non-equilibrium cells from their initial densities and temperatures. This algorithm could potentially also be used to create a hybrid method between \grackle{} and ML, where \grackle{} handles the equilibrium cells, where it clearly outperforms ML, while ML tackles the non-equilibrium cells that \grackle{} spends a significant time on. 

\begin{table}[]
    \centering
    \begin{tabular}{c|cc|c}
        Parameter & Min & Max & Unit \\\hline
        $\rho$ & -21 & -19 & log(g/cm$^3$) \\
        $\rho_\mathrm{H}$ & -21 & -19 & log(g/cm$^3$)\\
        $\rho_\mathrm{H^+}$ & -27 & -25 & log(g/cm$^3$)\\
        $\rho_\mathrm{H^-}$ & -36 & -31 &log(g/cm$^3$)\\
        $\rho_\mathrm{H_2}$ & -33 & -21 & log(g/cm$^3$)\\
        $\rho_\mathrm{H_2^+}$ & -35 & -29 & log(g/cm$^3$)\\
        $\rho_\mathrm{He}$ & -21 & -19 & log(g/cm$^3$)\\
        $\rho_\mathrm{He^+}$ & -33 & -29 & log(g/cm$^3$)\\
        $\rho_\mathrm{He^{++}}$ & -43 & -40 & log(g/cm$^3$)\\
        $\rho_\mathrm{e^-}$ & -28 & -25 & log(g/cm$^3$)\\
        $E$ & 12 & 13 & log(erg/g)\\
        $t_f$ & -2 & 6 & log(yr)
    \end{tabular}
    \caption{High density/energy ranges using the same units as Table \ref{tab:ranges}. These ranges are used to capture the parameter space where \grackle{} has a lower performance.}
    \label{tab:highdens}
\end{table}

\subsection{Iteration} \label{sec:iteration}
\begin{figure}[]
    \centering
    \includegraphics[width=0.4\textwidth]{Figures/legend.pdf}
    \includegraphics[width=0.5\textwidth]{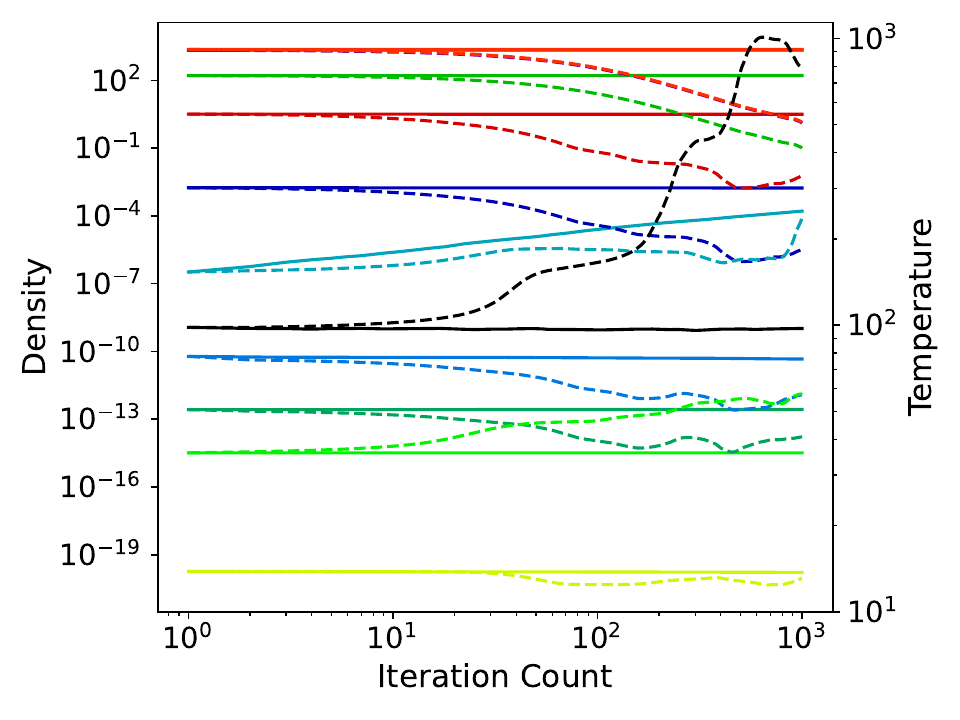}
    \includegraphics[width=0.5\textwidth]{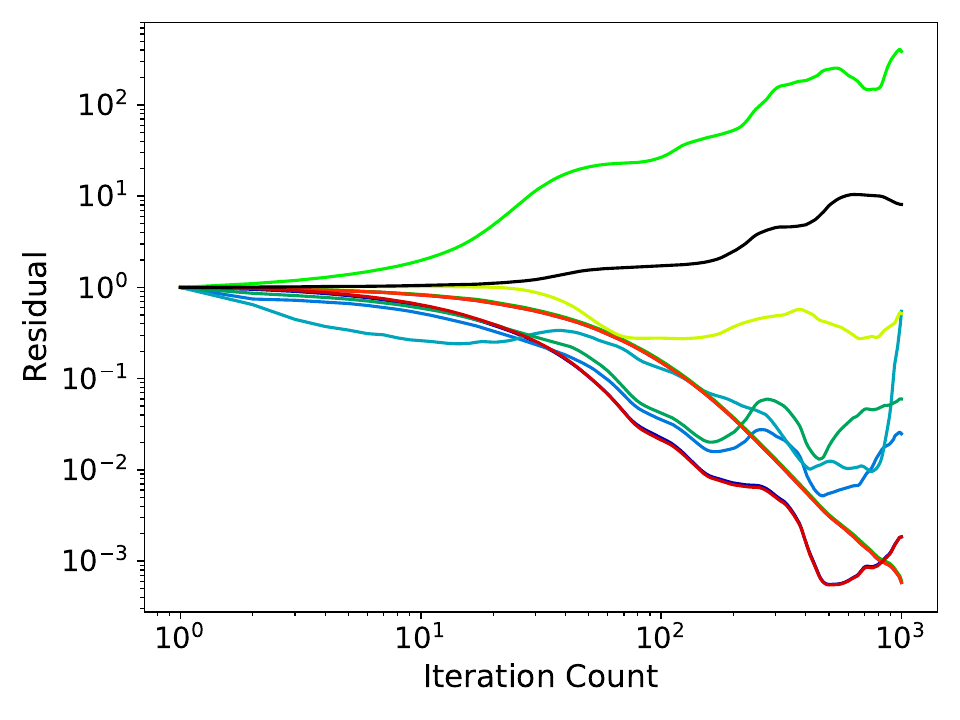}
    \caption{Testing the stability of ML predictions over repeated iterations. Top: A comparison of the outputs of \grackle{} and the fiducial ML model as a function of iteration count. The solid lines represent values calculated by \grackle{}, while the dashed lines represent predicted values using the fiducial model.
    Bottom: The residual error between \grackle{} and the fiducial ML model for each species, as a function of iteration count.}
    \label{fig:iterations}
\end{figure}

\noindent So far, we have shown that predicting the output of \grackle{} works for a single step in time. However, \grackle{} is used iteratively on each cell, to evolve the chemical abundances in lockstep with the hydrodynamical solver, while remaining numerically stable. 
To integrate an ML model into a hydrodynamical simulation code, it will need to be numerically stable over iterations, where the output of the previous step will be used as input for the following step. While the ML models predict \grackle{} extremely reliably, there remains a minor error between ML predictions and \grackle{} calculations (approximately 0.6 dex as noted in \S \ref{sec:onezone} and higher for rapidly evolving cells). However, this error may compound over several iterations. We perform a test, using the fiducial model, similar to the one-zone test described in the previous section, but instead of starting from the same initial cell with a different time step, we keep the time step constant and allow for iteration until a target time is reached. This is done both for \grackle{} and the ML model. \\
\indent To achieve this we select a cell with a small amount of variance, and use a constant time step of $1$ yr, up until a final time of $10^3$ yr, making for $10^3$ iterations. The resulting evolution is shown in Figure \ref{fig:iterations}. The top panel shows the evolution over the iterations, while the bottom panel shows the residual error between the ML prediction and \grackle{}. It's clear from this plot that the ML method does not align with \grackle{} for a repeated number of iterations. The discrepancy between the models worsens with the number of iterations, as is shown by the residuals pictured on the bottom plot of Figure \ref{fig:iterations}. The total density is not constant, failing a first test on convergence. This implies that the ML model is not stable over repeated iterations. One cause for this mismatch is that the output space of \grackle{} is greater than the input space for ML. \grackle{}, after evolving a cell, may calculate that one or several species densities have values outside of the ranges shown in Table \ref{tab:ranges}. The ML model learns this behaviour appropriately. The next iteration uses the outputs of the previous iteration as input, and since these values lie outside of the ranges shown in Table \ref{tab:ranges}, the ML model is forced to extrapolate. This significantly decreases the quality of the predictions. However, the initial cell used in Figure \ref{fig:iterations} was chosen such that the values predicted by \grackle{} would fall within the ranges of the training data shown in Table \ref{tab:ranges}. \\
\indent The first few iterations show no transgression into regions outside of these ranges, while \cite{branca} show that their trained model generalizes for density and temperature ranges at least half a dex outside of their chosen ranges for a single time step, implying a value slightly outside of these ranges should not cause a major error. The fact that the error is still present and grows with the iteration number while the input is not further than 0.5 dex outside of the ranges shown in Table \ref{tab:ranges}, implies that this effect is not relevant in the particular case of this cell, potentially pointing to a more fundamental cause for the growing discrepancy. \\
\indent One potential way to constrain the erroneous behaviour of the ML results may be to introduce charge and baryon conservation laws into the initial conditions and after prediction. These may constrain the resulting values, decreasing the rate of divergence. Future work will take this effect into account. However, there is also a more fundamental issue at play in this case. The DeepONet architecture only trains on a single time step, and the minor errors between \grackle{} and ML predictions compound with repeated application. This is a known issue for several neural network architectures \citep{stiasny_2021}, and methods to alleviate this issue for other high-level architectures already exist \citep[e.g][]{geneva_2020,list_2024} but these have not been applied to DeepONet in particular. Future work in integrating ML methods into hydrodynamical simulations will have to take this effect into account. Emulating a system of rate equations to work iteratively is however, outside the scope of this work. 

\section{Discussion and Conclusion} \label{Sec:Discussion}
\noindent Neural operators are machine learning techniques capable of learning solutions to ODE networks, such as the network of rate equations present in \grackle{}. We lay out a method for training these models, primarily based on the implementation of \cite{branca}, while also introducing three variant models (which we refer to as  \texttt{Wide}, \texttt{Deep} and \texttt{WideDeep}). We select initial conditions to span a wide range of densities and energies based on the densities and energies of a cosmological simulation, in order to train on typical values present in these simulations. \\
\indent We demonstrate that our ML models are able to match \grackle{} to an excellent degree of accuracy (see Figure \ref{fig:modelperformance}), and provide a computational speedup of up to a factor of 8 for certain density/energy ranges. 
Our relative errors between the ML result and the results from \grackle{} average in the range of $10^{-3}$, while \cite{branca} find relative errors around $10^{-2}$. However, they note an approximate speedup of 128, much greater than our results. Their comparison is however against \krome{} \citep{grassi2014}, with radiation effects taken into account. This increases the number of chemical reactions, increasing the complexity of the chemical model. Our simpler chemical network takes fewer computations until completion, making \grackle{} in this configuration faster than the network used by \cite{branca}. This increase in model complexity makes a direct comparison of the speedup challenging, and may also drive the difference in relative error. Nonetheless, our main results agree with those of \cite{branca} in that we both find excellent agreement between the ML approach and the direct chemistry solver for the one zone test and we both clearly show significant computational gains.

In terms of model variants we find that after training, the fiducial model performs well, with a typical cell having less than 0.6 dex error over the entire time range considered for the one-zone tests considered, however this may rise to over 1.5 dex for cells with higher variance. However, the \texttt{Wide} model appears to have the best overall performance, outperforming the fiducial model in terms of both speed and accuracy. In comparison the \texttt{Deep} and \texttt{WideDeep} models show larger relative errors (compared to \grackle), this may be explained by the amount of training, with the deeper models requiring more training for equal performance. 

\indent Overall, while these results show that DeepONet has the potential to work with cosmological simulations, it is still far from being able to fully replace the chemical network currently in place in many cosmological simulations. The reliability of these ML models to predict in lockstep with the hydrodynamical evolution of such a simulation is currently not at the level required for practical use. While for an individual time step the predictions are excellent, small differences expand rapidly with repeated iterations - quickly breaking numerical stability and accuracy. \\
\indent Another future challenge is the inclusion of GPU hardware in the cosmological simulation codes. The ML models laid out in this work perform best on GPU hardware, but many simulation codes exclusively use CPU hardware, limiting the potential speedup of ML methods. 
The training of ML models also benefits from GPU hardware, allowing larger networks to train faster on larger data sets, improving accuracy and performance. 
A full integration of ML methods into cosmological hydrodynamics simulations has great potential, but will have to overcome these challenges.

\section{Acknowledgements}
\noindent PvdB \& JR acknowledges support from Research Ireland via the Laureate programme under grant number IRCLA/2022/1165. JB \& JR acknowledges support from the Royal Society and Research Ireland under grant number URF\textbackslash R1\textbackslash 191132.  

\bibliographystyle{mn2e}
\bibliography{main}

\appendix
\section{Accuracy Plots} \label{sec:appendixAccuracy}
\begin{figure}[H]
    \centering
    \includegraphics[width=0.3\linewidth]{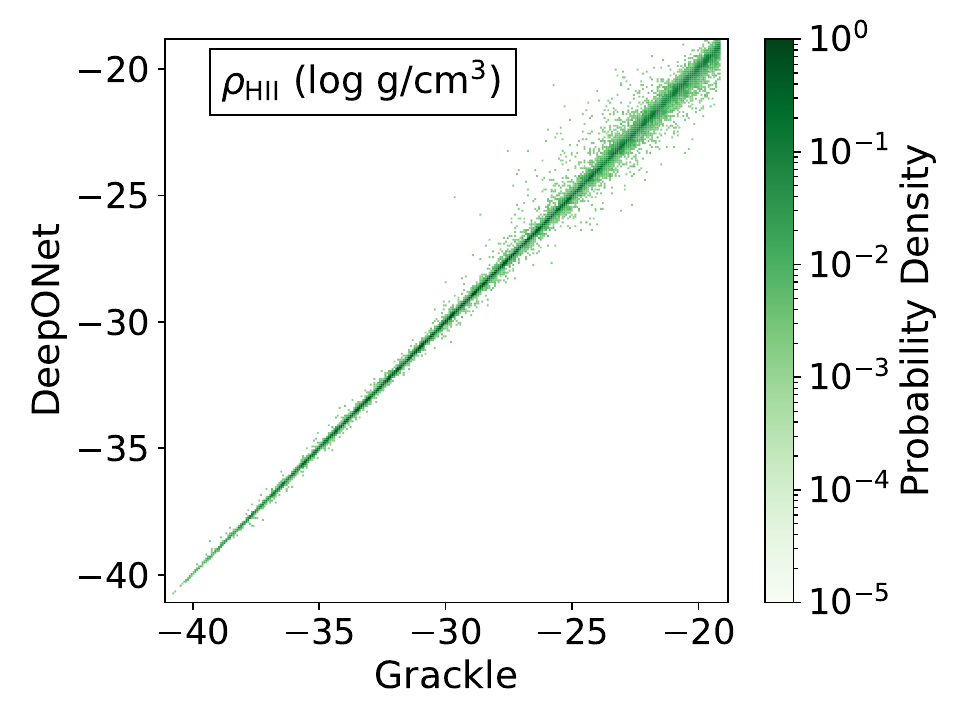}
    \includegraphics[width=0.3\linewidth]{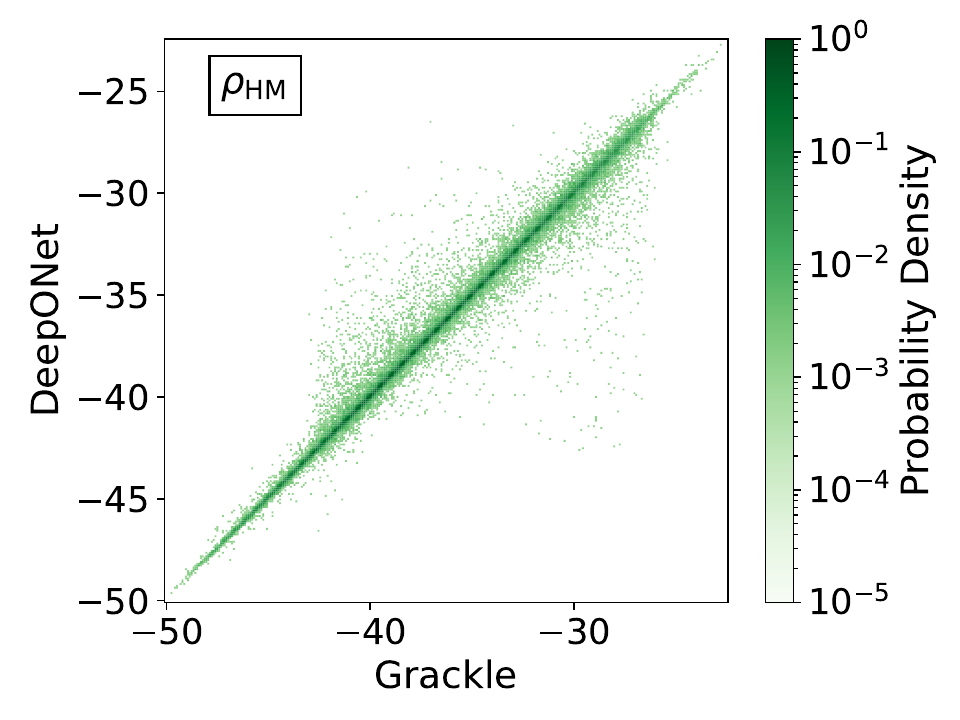}
    \includegraphics[width=0.3\linewidth]{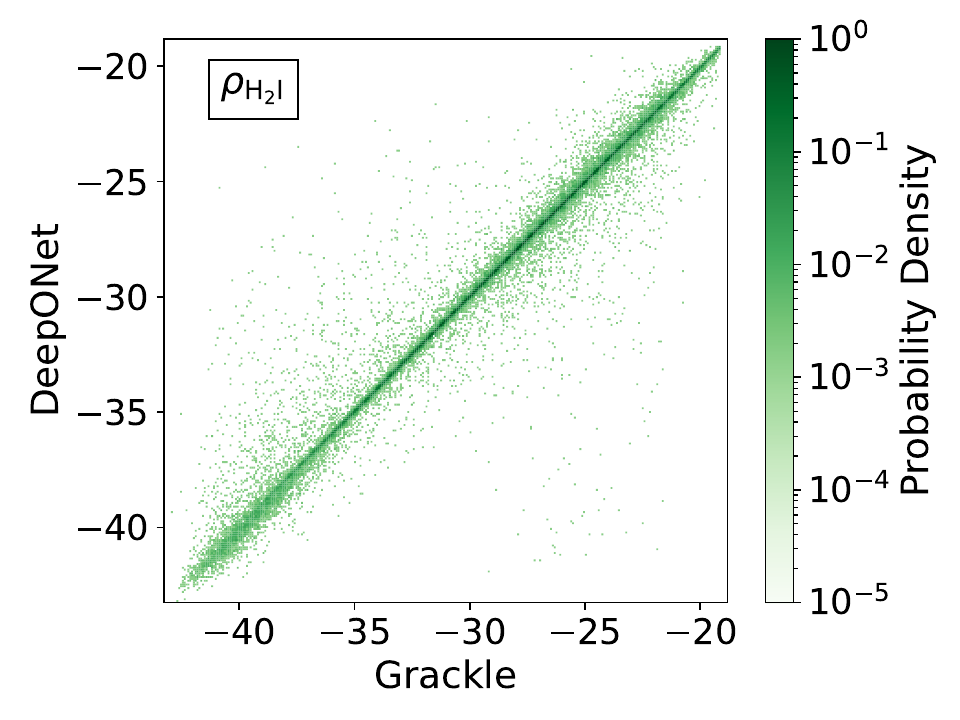}
    \includegraphics[width=0.3\linewidth]{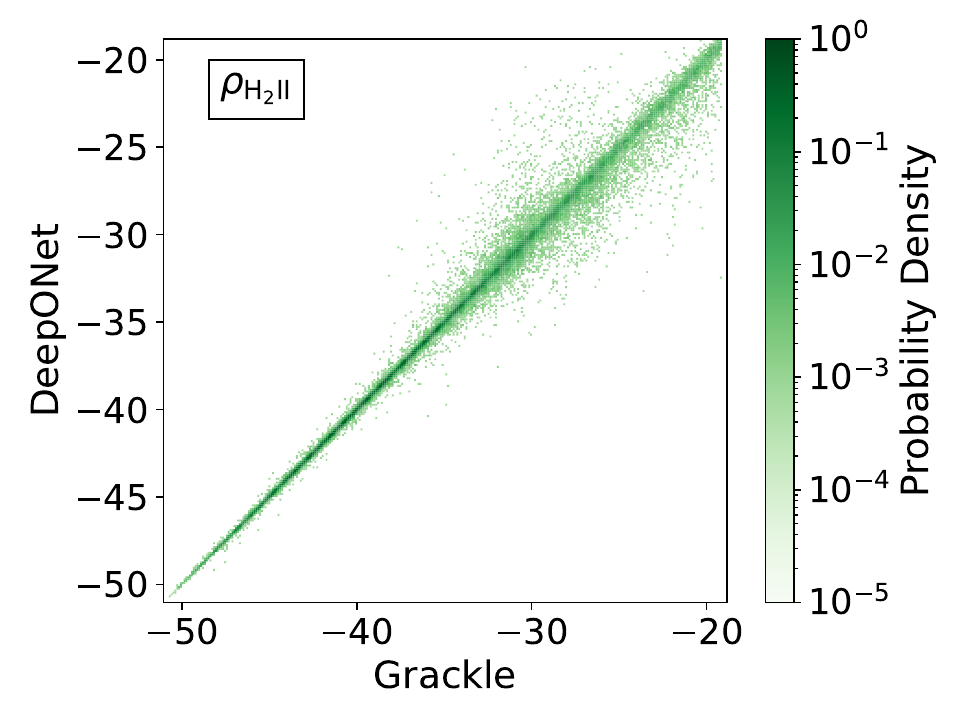}
    \includegraphics[width=0.3\linewidth]{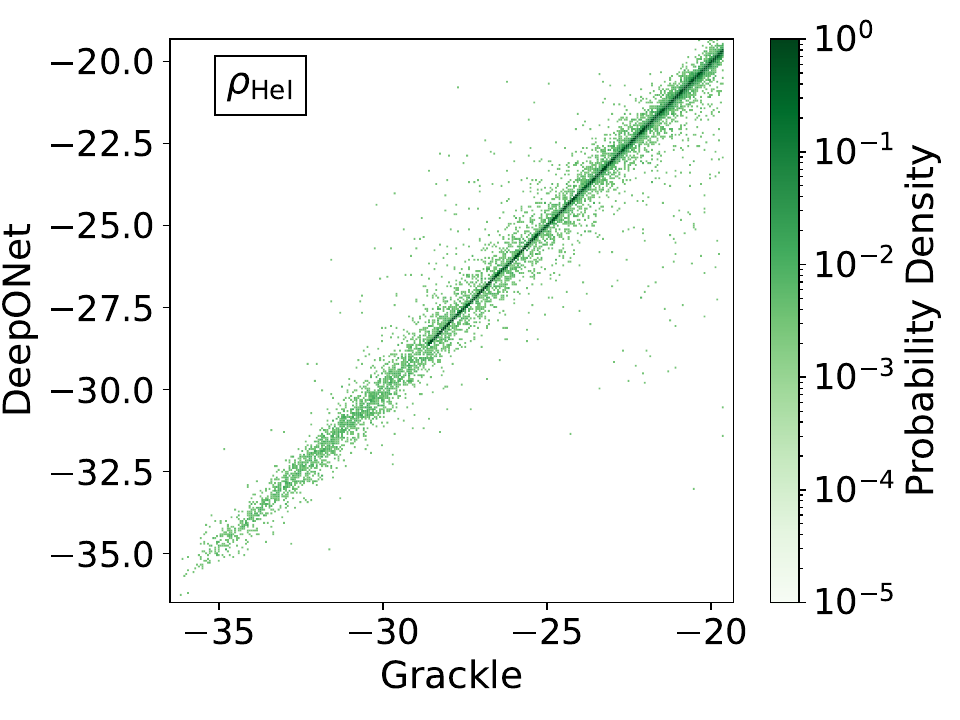}
    \includegraphics[width=0.3\linewidth]{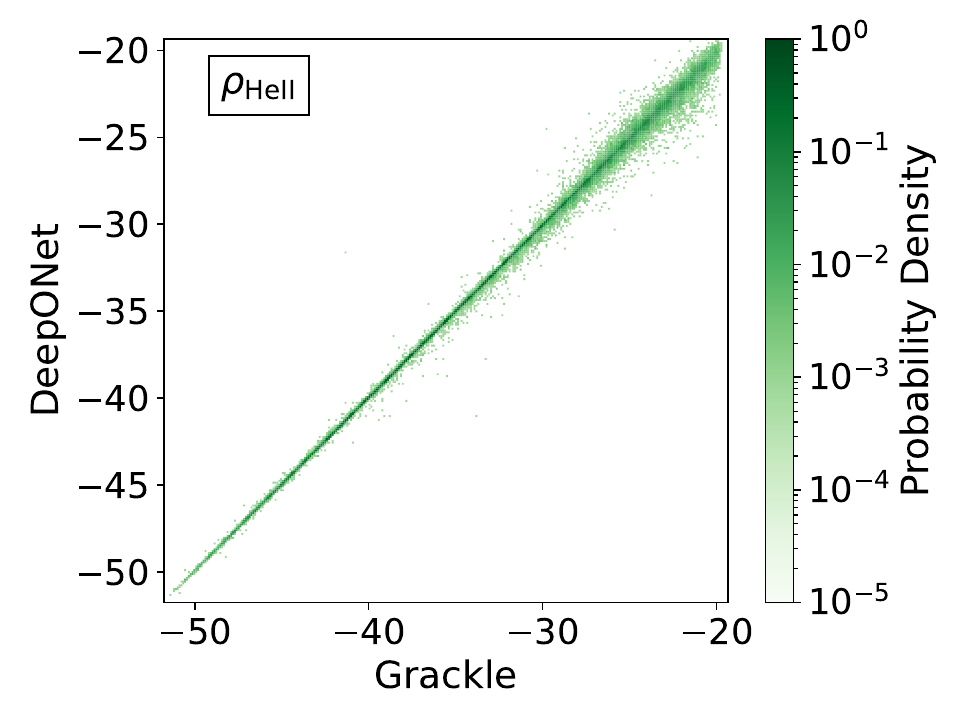}
    \includegraphics[width=0.3\linewidth]{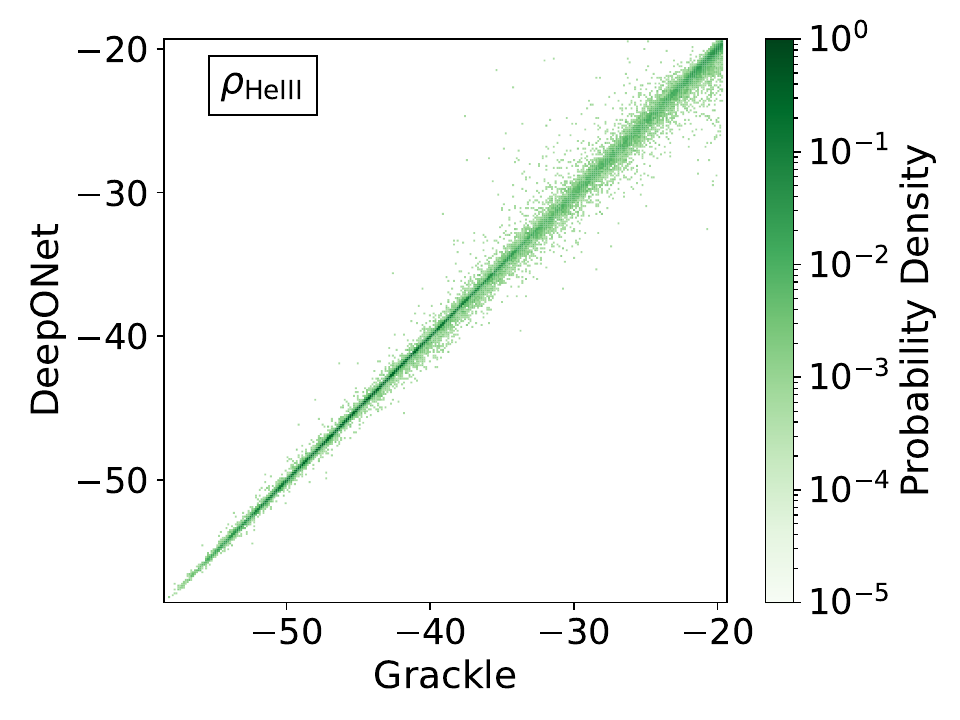}
    \includegraphics[width=0.3\linewidth]{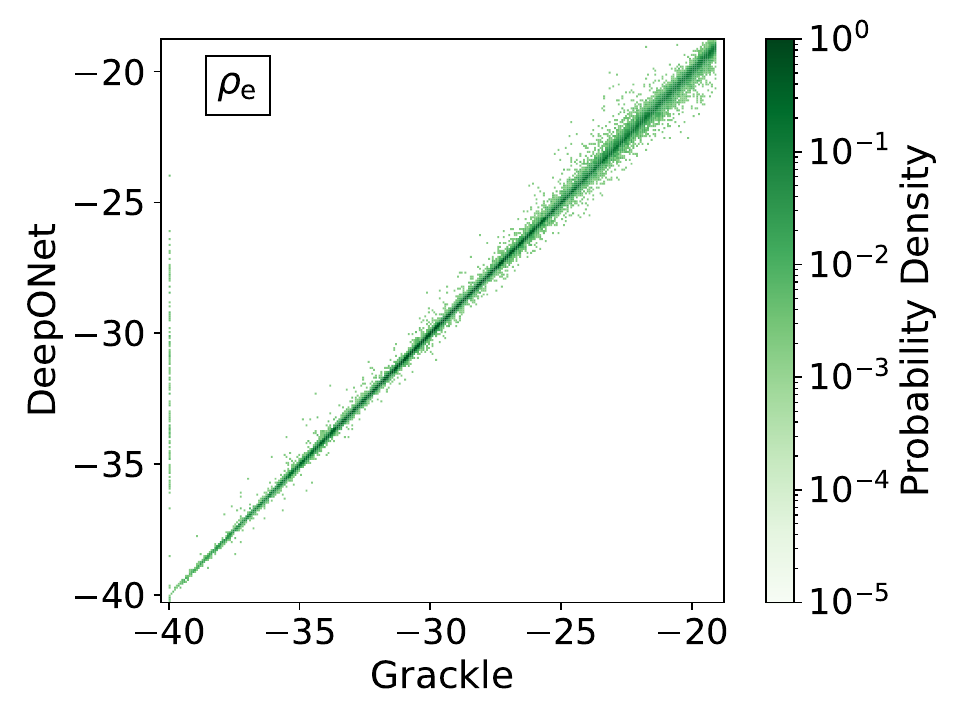}
    \includegraphics[width=0.3\linewidth]{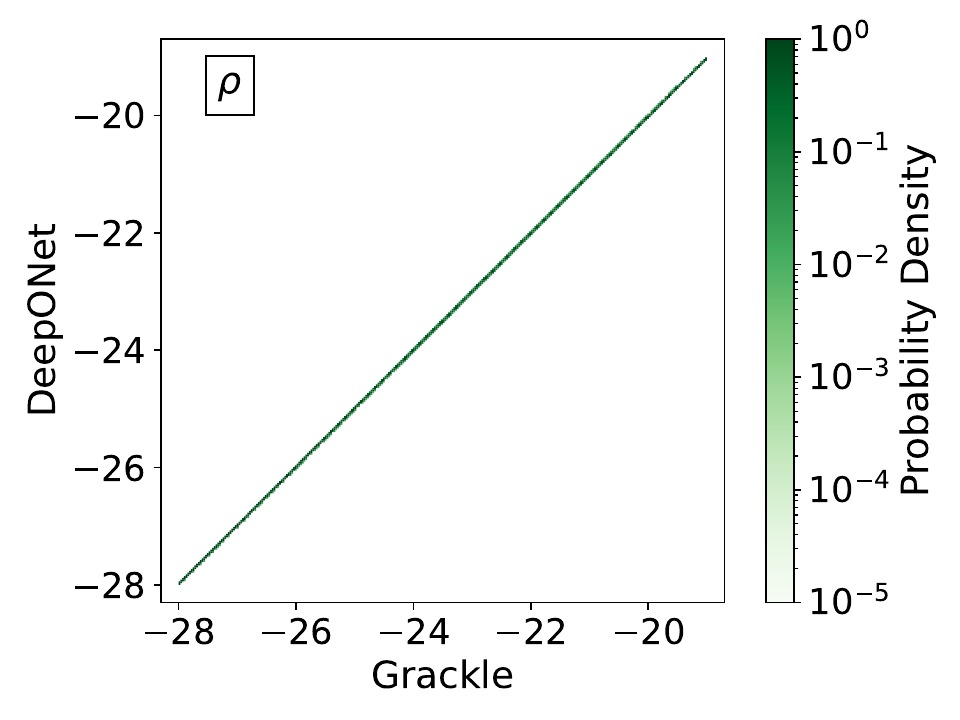}
    \includegraphics[width=0.3\linewidth]{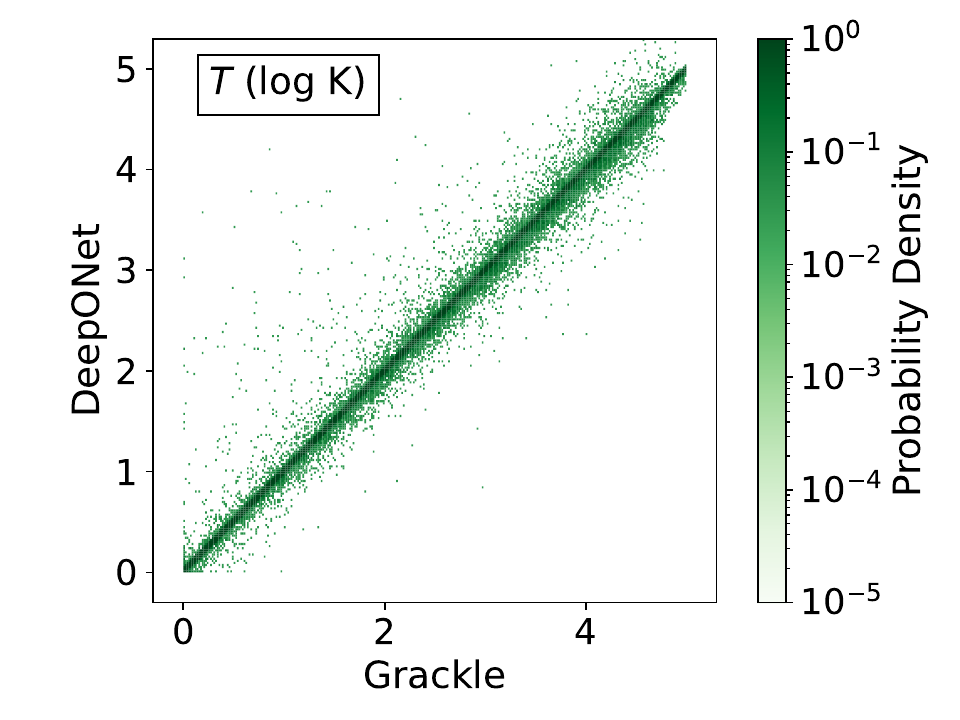}
    \caption{Probability distributions of values calculated with \grackle{} compared to predictions by the fiducial model for the species not shown in Figure \ref{fig:accuracy}, as well as the total density and temperature. A diagonal line represents an ideal match between \grackle{} and the fiducial model.}
    \label{fig:accuracyAppendix}
\end{figure}
\section{Relative Error Plots} \label{sec:appendixRelErr}
\begin{figure}[H]
    \centering
    \includegraphics[width=0.3\linewidth]{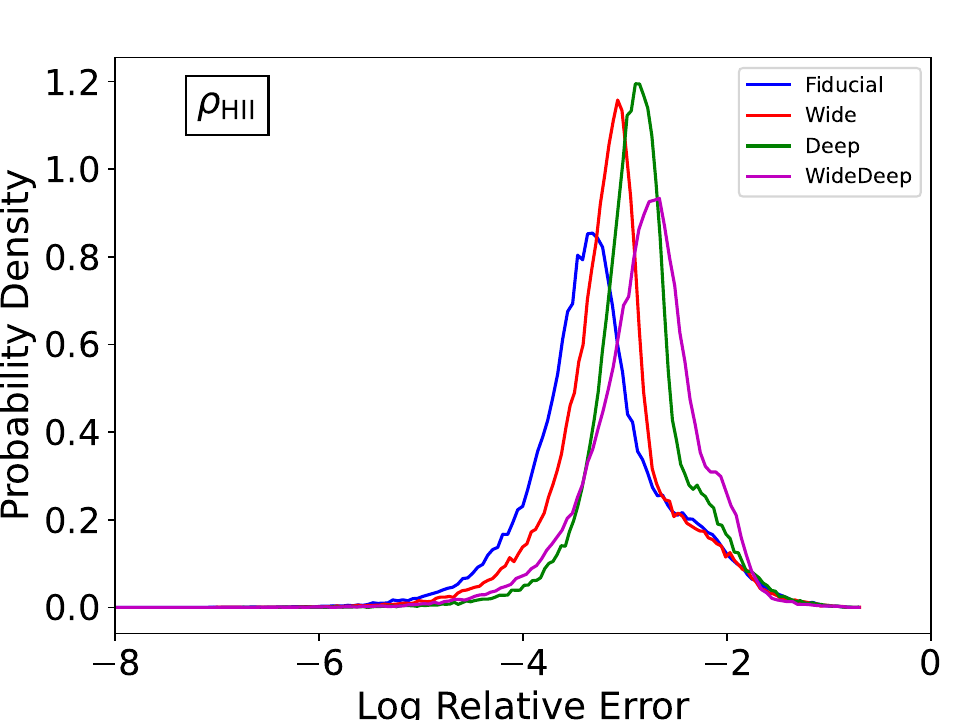}
    \includegraphics[width=0.3\linewidth]{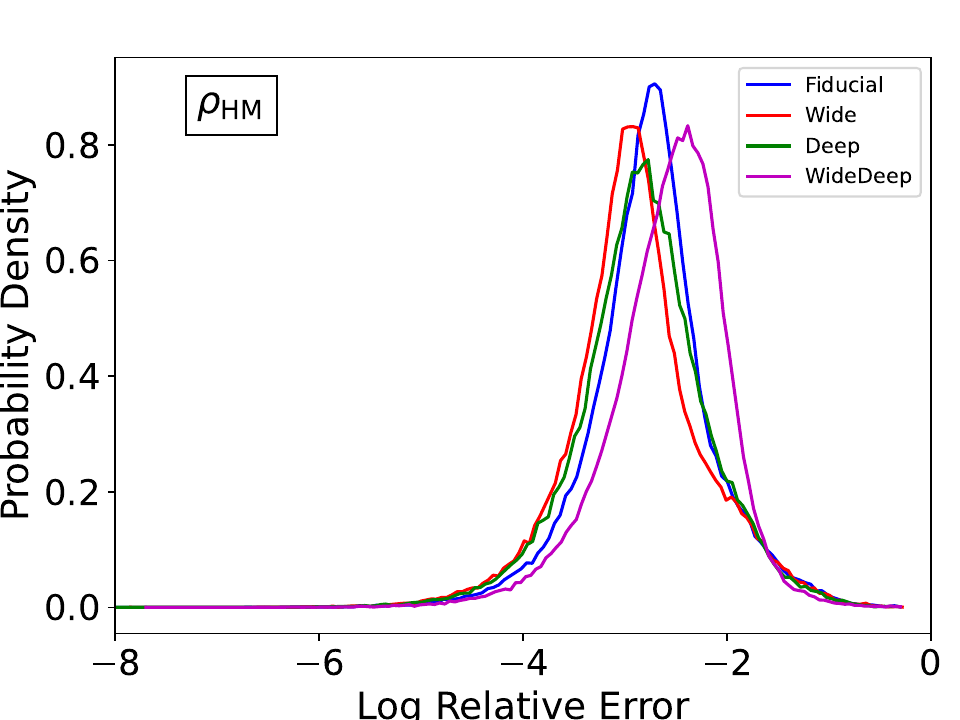}
    \includegraphics[width=0.3\linewidth]{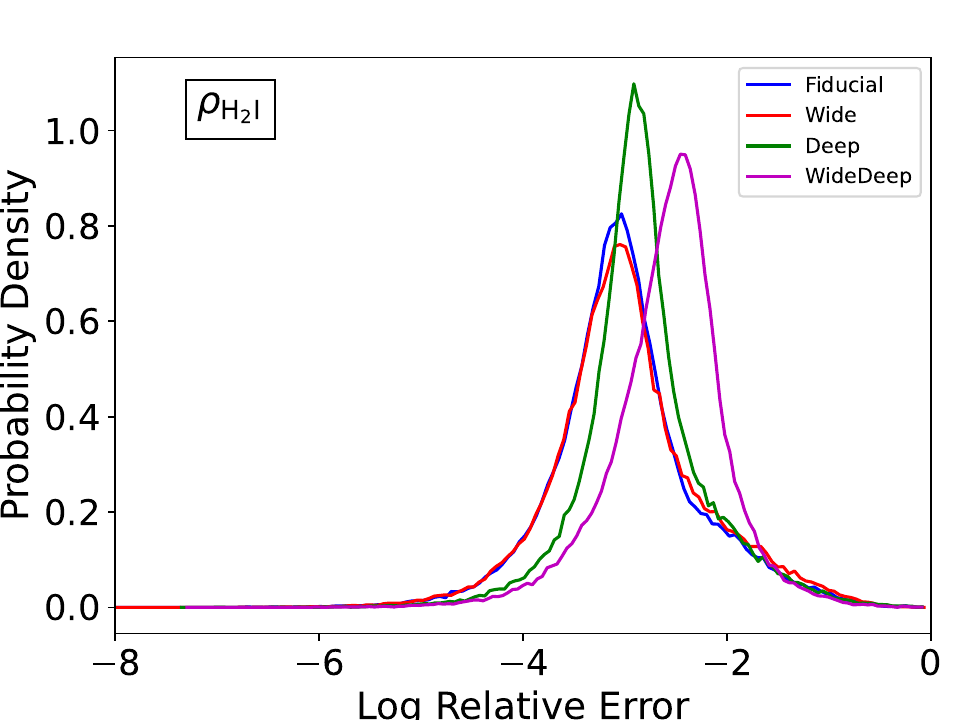}
    \includegraphics[width=0.3\linewidth]{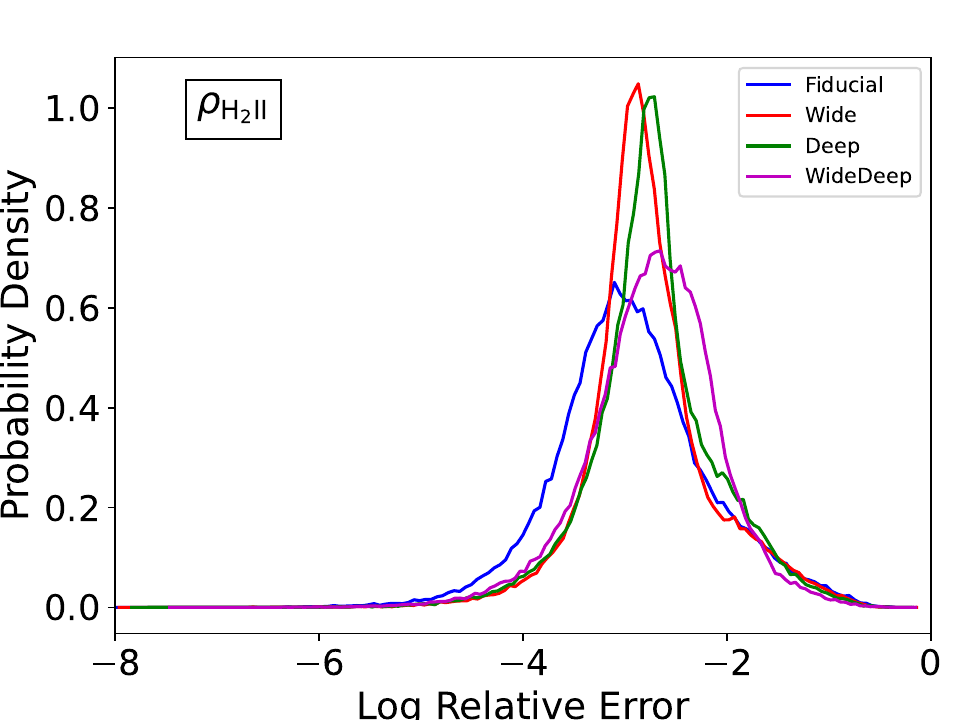}
    \includegraphics[width=0.3\linewidth]{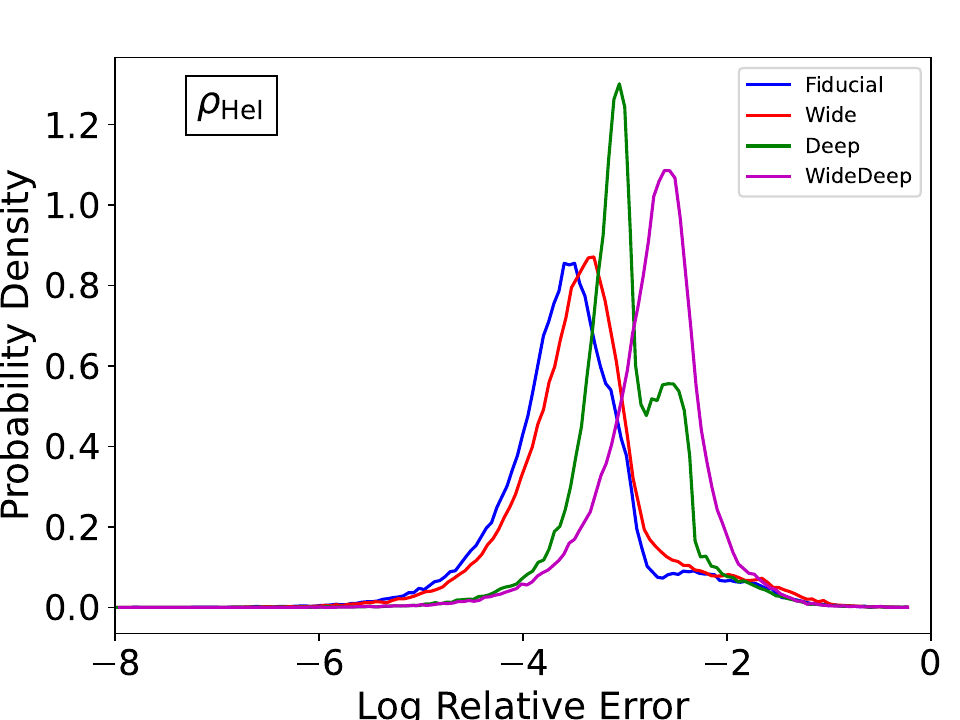}
    \includegraphics[width=0.3\linewidth]{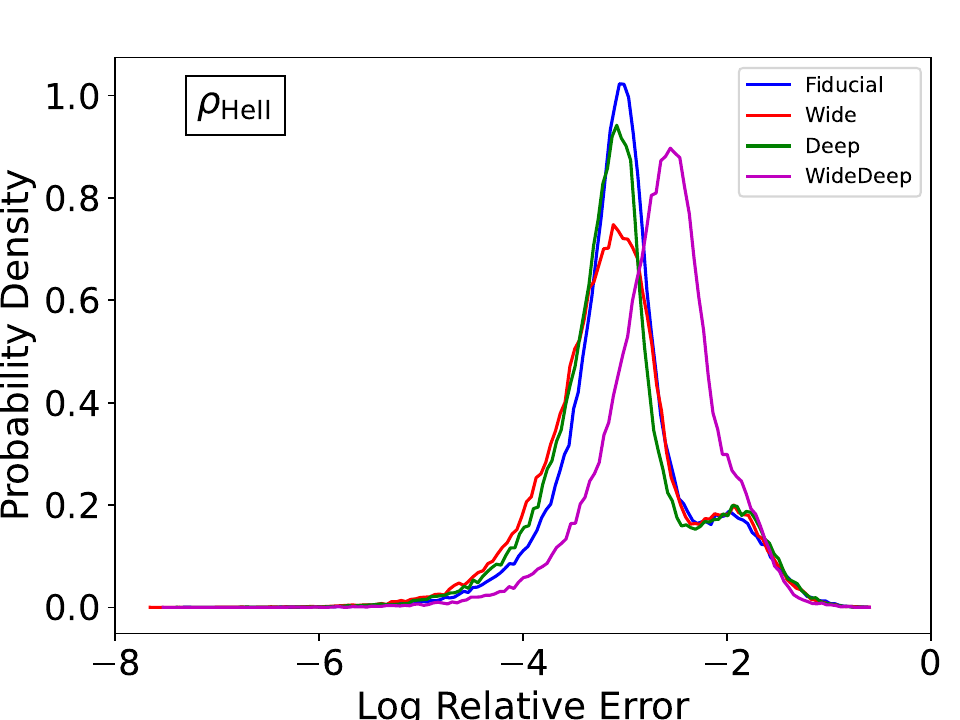}
    \includegraphics[width=0.3\linewidth]{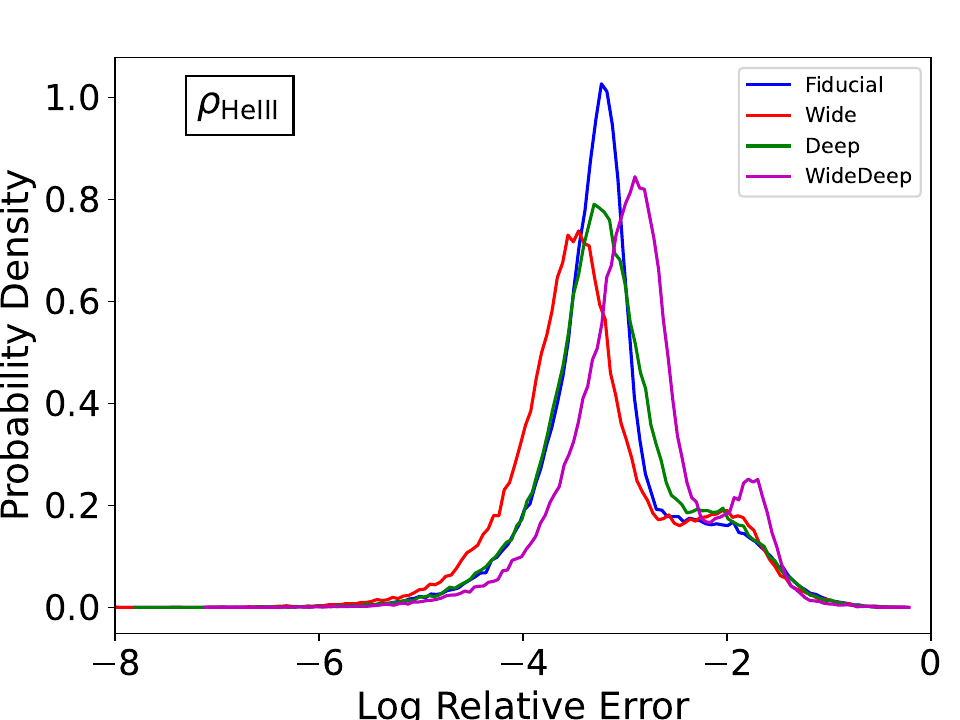}
    \includegraphics[width=0.3\linewidth]{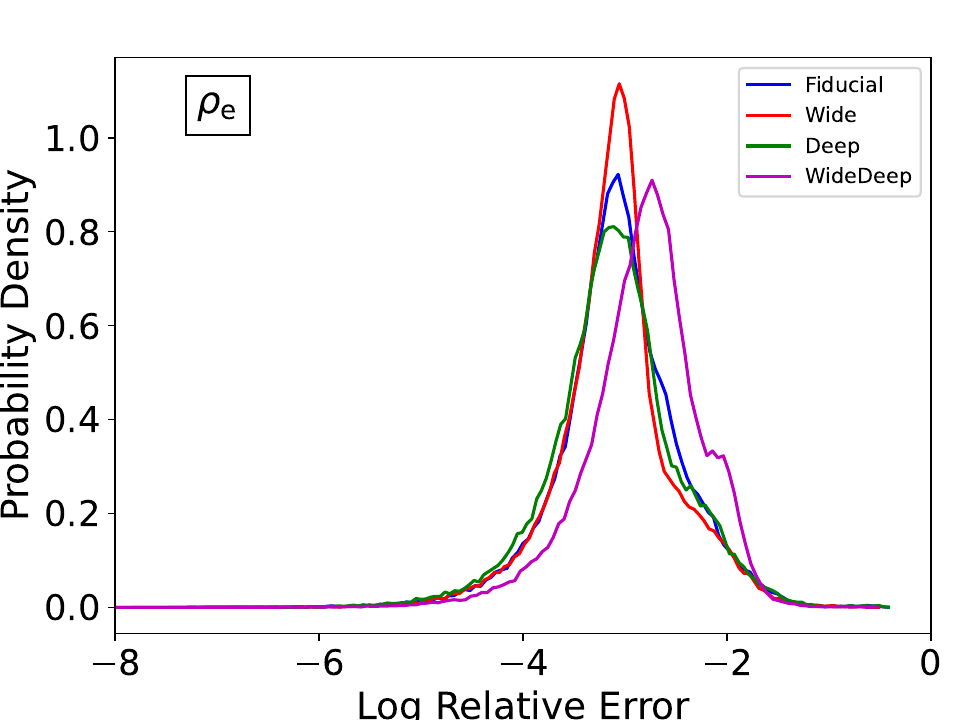}
    \includegraphics[width=0.3\linewidth]{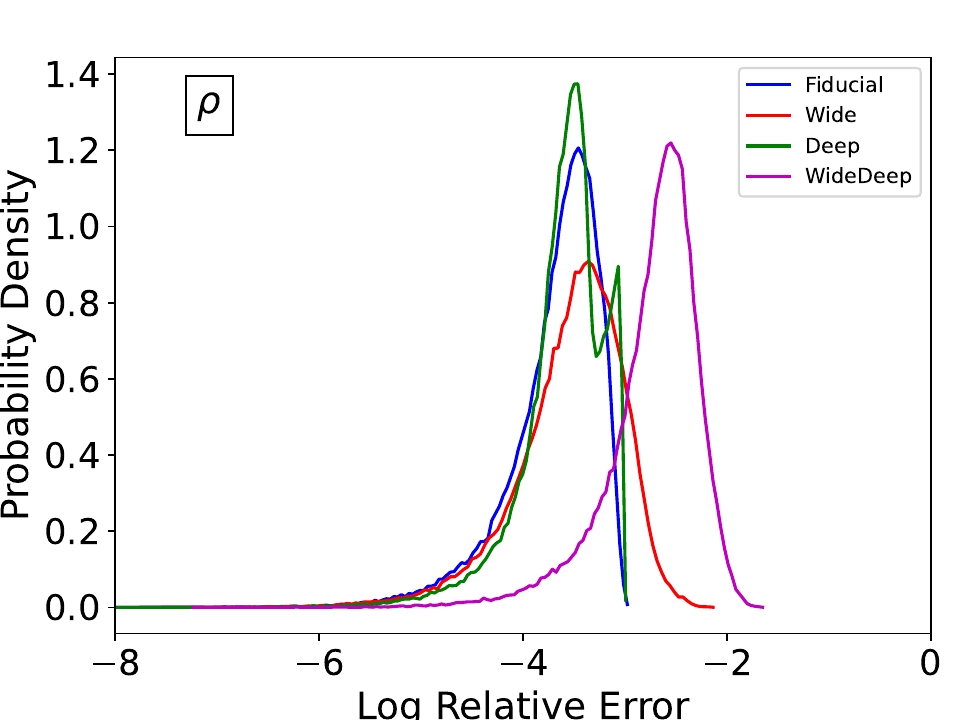}
    \includegraphics[width=0.3\linewidth]{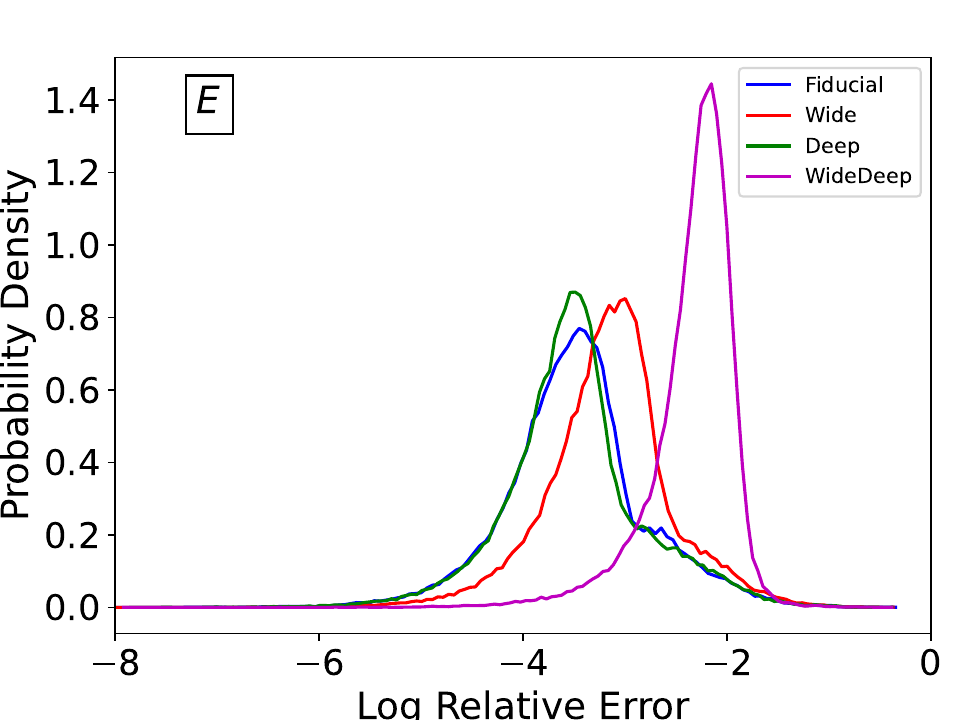}
    \caption{Probability distributions of relative errors between the four ML models and \grackle{} for the species not shown in Figure \ref{fig:modelperformance}, as well as the total density and internal energy. These values are calculated using equation \ref{eq:relerr}, based on Grackle and ML predictions of $10^6$ samples.}
    \label{fig:predictionDistributions}
\end{figure}


\end{document}